 \documentclass[11pt]{article}

\usepackage{bm}
\usepackage{graphicx}

\usepackage{hyperref}
\usepackage[numbers]{natbib}
\makeatletter 
\renewcommand\@biblabel[1]{#1.} 
\makeatother %

\newcommand{\be}{\begin{eqnarray}}
\newcommand{\ee}{\end{eqnarray}}

\newcommand{\cif}{\mbox{CIF}}

\title{High-Dimensional Variable Selection and Prediction under Competing Risks with Application to SEER-Medicare Linked Data}

\author{Jiayi Hou\textsuperscript{a}, Anthony Paravati\textsuperscript{b}, Ronghui Xu\textsuperscript{c*} and James Murphy\textsuperscript{b}}

\newcommand{\Addresses}{{
  \medskip
  \footnotesize

\textsuperscript{a} \textit{Altman Clinical and Translational Research Institute, University of California, San Diego,  CA 92093, U.S.A}\\
\textsuperscript{b} \textit{Department of Radiation Medicine and Applied Sciences, University of California, San Diego, CA 92093, U.S.A}\\
\textsuperscript{c} \textit{Department  of Family  Medicine and Public Health and Department of Mathematics, University of California, San Diego, CA 92093, U.S.A}\\
\textsuperscript{*} \textit{Corresponding author}
}}

\date{}

\begin{document}
\maketitle
\Addresses






\begin{abstract}
Competing risk analysis considers event times due to multiple causes, or of more than one event types. 
Commonly used regression models for such data include 1) cause-specific hazards model, which focuses on modeling one type of event while acknowledging other event types simultaneously; and 2) subdistribution hazards model, which links the covariate effects directly to the cumulative incidence function. 
Their use and in particular statistical  properties in the presence of high-dimensional predictors are largely unexplored. Motivated by an analysis using the linked SEER-Medicare database for the purposes of predicting cancer versus non-cancer mortality for patients with prostate cancer, we study the accuracy of prediction and variable selection of existing statistical learning methods under both models using extensive simulation experiments, including different approaches to choosing penalty parameters in each method. We then apply the optimal approaches to the analysis of the SEER-Medicare data. 
\end{abstract}



\section{Introduction}


As an illustration project of how information contained in patients' electronic medical records can be harvested for the purposes of precision medicine, we consider the large data set linking the
Surveillance, Epidemiology and End Results  (SEER) Program database of the National Cancer Institute with the federal health insurance program Medicare database for  prostate cancer patients of  age 65 or older.  
Each year 180,000 men are diagnosed with prostate cancer in the US, and the important clinical decision commonly encountered in this patient population is whether to pursue aggressive cancer-directed therapy in the presence of pre-existing comorbidities. Prostate cancer can progress slowly, and a proportion of men will die of competing causes before their prostate cancer becomes symptomatic. Current clinical guidelines for the management of prostate cancer instruct clinicians to make treatment decisions based on two factors: 1) an estimation of the aggressiveness of a patient's tumor; and 2) estimation of a patient's overall life expectancy \cite{thompson2007guideline}. 
Classical cancer-specific survival prediction relies on three main risk factors: tumor stage, Gleason score, and prostate specific antigen (PSA). On the other hand, currently no tool exists to predict non-cancer survival for this patient population.
As non-cancer and cancer survival are not independent, i.e.~so-called competing risks in the statistical literature, an accurate comprehensive survival prediction tool should consider both types of risks simultaneously.  

Competing risks occur when multiple types of failures co-exist and the occurrence of one type of failure may prevent the observation of the other types of failure. In addition the failure times may be subject to right-censoring. 
In the regression settings the  Cox proportional hazards model can be used
to model the so-called cause-specific hazards, and existing software for fitting the Cox model for classical survival data without competing risks can be used to fit the proportional cause-specific hazards model \citep{kalbfleisch2011statistical}.
Under this model, however,  the dependence of the cumulative incidence function of a particular failure type on the covariates involves  
also the effects of the covariates on the cause-specific hazards of all other types of failures.
 Beyersmann \textit{et al.}~\cite{beyersmann2007competing} showed as an example  in patients receiving peripheral blood stem-cell transplantation, while the cause-specific hazard ratio for certain baseline risk factors of bloodstream infection (competing with the event of neutropenia) might be similar, the corresponding cumulative incidence functions can be quite different.

In order to link the covariates directly to the cumulative incidence functions (CIF), 
 Fine and Gray \cite{fine1999proportional} proposed to model the subdistribution hazards.
 The proportional hazards modeling of the subdistribution hazards, also known as Fine-Gray model, has gained  popularity in recent years. 
 The proportional cause-specific hazards model, and the proportional subdistribution hazards model are typically not valid at the same time, and limited empirical experiences seem to indicate that in real data applications the two models can lead to similar conclusions \cite[]{geskus}. 

We restrict our analysis to patients diagnosed between 2004 and 2009 in the SEER-Medicare database. 
After excluding additional patients with missing clinical records, we have a total of 57,011 patients who have information available on 7 relevant clinical variables (age, PSA, Gleason score, AJCC stage, and AJCC stage T, N, M, respectively), 5 demographical variables (race, marital status, metro, registry and year of diagnosis), plus 8971 binary insurance claim codes. 
We assumed that the survival prediction would occur at the time of diagnosis, therefore we used clinical and demographic information at the time of diagnosis, and insurance claims data during the year prior to diagnosis. Insurance claims capture medical diagnoses and procedures through HCPCS codes, ICD-9 diagnosis codes, and ICD-9 procedure codes. These claims indirectly describe events that occur in surgical procedures, hospitalization and outpatient activities. We converted each unique insurance claim code into a binary variable denoted 1 if the claim appeared anytime in the year before diagnosis, and 0 if the code was absent. 
Until December 2013 (end of follow-up for this data) there were a total of 1,247  deaths due to cancer, and 5,221 deaths unrelated to cancer. It is  well understood that for time-to-event data, the number of events dictates the effective sample size, so in this case we have more predictors to consider than the effective sample size.

Researchers have studied different approaches  to analyze survival data with high dimensional covariates. Notably, Tibshirani \cite{tibshirani1997lasso} proposed the least absolute shrinkage and selection operator (LASSO) under the Cox proportional hazards model.  Zhang and Lu \cite{zhang2007adaptive} investigated the statistical properties of adaptive LASSO for the Cox proportional hazards model. Hothorn \textit{et al.}~\cite{hothorn2006survival} introduced a random forest algorithm and a generic gradient boosting algorithm for right-censoring data. When considering  theoretical aspects, Bradic \textit{et al.}~\cite{bradic2011regularization} studied a group of penalty functions and established strong oracle properties of non-concave penalized methods for ultra high dimensional covariates in the presence of right-censoring.  In comparison, very few high-dimensional methods have  been developed in the presence of competing risks. 
Binder \textit{et al.}~\cite{binder2009boosting} first proposed a boosting approach for fitting the proportional subdistribution hazards model. 
Very recently (published online at the time of submission of this manuscript) Fu \textit{et al.}~\cite{fu:etal:2016} considered penalized approaches under the same model.
Given the high-dimensional nature of our data, in this paper we will consider both the proportional cause-specific hazards (PCSH) model and the proportional subdistribution hazards (PSDH) model, and investigate 
the accuracy of variable selection and prediction using existing computational software under either model.
This leads to the Binder \textit{et al.} approach under the PSDH model, and LASSO approach under the PCSH model, both being readily implemented and applicable to our large data set. 
Both approaches rely critically on the selection of a `penalty' parameter, and there are different ways to select this parameter. We will empirically evaluate these different methods using  Monte Carlo simulations. 
 The ultimate goal is to assess the prediction accuracy of the cumulative incidence function as a risk assessment tool.  

The remainder of this paper is organized as follows: in Section \ref{competing risk model}, we review the proportional cause-specific hazards model and the proportional subdistribution hazards model. In Section \ref{variable selection}, we review the relevant statistical learning methods that have been or  can be feasibly implemented to analyze competing risks data under each model. In Section \ref{Simulations}, we conduct comprehensive simulation studies on these methods  with varying numbers of predictors (relative to the sample size), 
that are continuous or binary (and in case of binary, sparse or not sparse). In Section \ref{realdata}, we apply the statistical learning methods under either model to classify prostate patients from the SEER-Medicare linked data into different risk groups according to their predicted cumulative incidence functions. 
Finally,  Section \ref{Discussion} contains discussion and directions for future work.

\section{Competing Risk Models}\label{competing risk model} 

Let $\epsilon=1, ..., J$ be the cause or type (we use the two words interchangeably in the following) of failure. 
Let $T = \min_{j=1}^J \tilde{T}_{j} $ denote the observed failure time if there is no censoring which is due to one of the causes, while failures from other types or causes are latent. 
Let $X_i = \min(T_i, C_i)$, $\delta_i = I(T_i \leq C_i)$,  where $C_i$ is the potential censoring time, and is assumed non-informative.
Denote $S(t)= P(T>t)$ the survival function of $T$.
The cumulative incidence function (CIF) for failure type $j$ is $F_j(t) = P(T\leq t, \epsilon = j)$. Obviously $S(t)= 1- \sum_{j=1}^J F_j(t)$, and $\sum_{j=1}^J F_j(\infty) =1$.
Denote the cause-specific hazard function of type $j$ as $\lambda_j(t) = \lim_{\Delta t \rightarrow 0+} Pr(t \le T < t+\Delta t, J=j | T \ge t) / \Delta t$. Then one can also show that 
\begin{eqnarray}
\label{CIF}
F_{j}(t) = \int_{0}^{t}\lambda_{j}(u)S(u)du, 
\end{eqnarray} 
leading to a nonparametric estimate of the CIF if we use
 \cite{fleming2011counting}: 
\begin{eqnarray}
\hat{\lambda}_j(t_i) &=&\frac{d_{ji}}{n_{i}}
\ee
where $d_{ji}$ denotes the number of  failures from cause $j$ at time $t_{i}$  and  ${n_{i}}$  the number of subjects at risk at $t_{i}$, and 
\be
\hat{S}(t)& =& \prod_{j:t_{i}\le t} \bigg(1-\sum_{j=1}^{J}\hat{\lambda}_{j}(t_{i})\bigg).
\end{eqnarray}
Then we have $\hat{F}_{j}(t)= \sum_{i: t_{i} \le t} \hat{p}_{j}(t_{i})$, where 
$\hat{p}_{j}(t_{i}) = \hat{\lambda}_{j}(t_{i}) \hat{S} (t_i^-)$.  
While  $\hat{F}_{j}(t)$ is a complex function of the $\hat{\lambda}_j(t_i) $'s,   the $(1-\alpha)100\%$ pointwise (for each $t$) confidence intervals can calculated using the `Cuminc()' function in the R package `mstate'.

\subsection{The PCSH Model} 

Given a vector of covariates $Z$, under the proportional hazards assumption of the cause-specific hazard function we have
\be
\lambda_j(t| Z) = \lambda_{0j}(t) \exp(\beta_j' Z),
\ee
for $j=1, ..., J$. 
To estimate $ \beta_j$, we can use any software for the regular Cox model by treating all other types of events as if censored, one type of events at a time.
This is because the (partial) likelihood for all event types factors into a separate
likelihood function for each event type, and the likelihood function for each event type treats all other types of events as if censored. 

To estimate the cumulative incidence function given $Z=z_0$, we have similar to the above nonparametric estimation:
\begin{eqnarray}
\label{parametricCIF}
\hat{F}_{j}(t) 
&=&\int_{0}^{t}\hat{S}(u; z_{0})d\hat{\Lambda}_{j}(u; z_{0}) \\ \nonumber
                        & = &  \sum_{i=1}^{n} \frac {\hat{S}(X_{i}; z_{0})\delta_{ji}I(X_{i} \le t)\exp(\hat{\beta}_{j}' z_{0})} {\sum_{i^{\prime}=1}^{n}I( X_{i} \le X_{i^{\prime}})\exp({\hat{\beta}_{j}' Z_{i^{\prime}}})},          
                        \end{eqnarray}
        where  $\hat{S}(u; z_{0})= \exp \{-\sum_{j=1}^{J}\hat{\Lambda}_{j}(u;z_{0}) \}$, $\hat{\Lambda}_{j}(u;z_{0}) = \hat{\Lambda}_{0j}(u)\exp(\hat{\beta_{j}}' z_{0})$, and the baseline cumulative hazard $\hat{\Lambda}_{0j}(u)$ is a Breslow-type estimator \citep{breslow1974covariance}.                Notice that in estimating the overall survival function $\hat{S}$ we need to fit the models for all event types, even if we are only interested in the CIF of type $j$.
        
        A $(1-\alpha)100\%$ pointwise confidence interval can be computed following Cheng \textit{et al.}~\citep{cheng1998prediction}.
We implemented an R package `CompetingRisk' \cite[]{RpkgCompetingRisk}  to compute the above estimator of the CIF with its pointwise confidence intervals.

\subsection{The PSDH Model} 

Gray \cite{gray1988class} introduced the subdistribution hazard function as
$\tilde\lambda_j(t) = -\frac{d}{dt} \log\{1- F_j(t) \}$.
Under the proportional hazards  assumption of the subdistribution hazard function for cause 1 we have \cite{fine1999proportional}
\be\label{eq:psdh}
\tilde\lambda_1 (t|Z) = \tilde\lambda_{0}(t) \exp (\beta' Z).
\ee
It is easy to see that model (\ref{eq:psdh})
 provides a direct way  to estimate the CIF of cause 1, 
 so that there is no need to fit models for the other causes in order to estimate $\mbox{CIF}_1$. 
 
 Fine and Gray \cite{fine1999proportional} proposed estimating equations for $\beta$. Geskus \cite[]{geskus2011} further showed that these estimating equations can be  solved using weighted Cox regression, i.e.~software for the regular Cox model incorporating weights. 
 The baseline subdistribution hazard  is again estimated using a modified version of Breslow's estimator. The $(1- \alpha)100\%$ pointwise confidence intervals can be constructed by sampling standard normal random variables and otherwise closed-form formulas \cite[]{fine1999proportional}.

\section{Regularization} \label{variable selection} 

Classical statistical methods,  such as stepwise regression, 
have been known to suffer from inconsistency and are computationally infeasible  when the number of covariates is equal to or greater than the (effective) sample size. 
A group of statistical learning methods, in particular supervised learning has shown good performance empirically 
when the data is of high-dimensionality \cite[]{fan2001scad}. The  goals of these methods are \cite[]{buehlmann:sara:book} 1) prediction: to find a set of covariates which results in minimal prediction error in independent test data; 2) variable selection: estimate the true sparsity pattern with low false positive rate for each covariate. 
In theory, consistent variable selection requires stronger assumptions, which are more difficult to meet in practice. 
Fortunately for our application prediction is of interest, and in this paper we will  study the  performance of statistical learning methods in estimating the true cumulative incidence function $F_{j}$. 
These statistical learning methods  often involve the selection of a  tuning parameter, 
based on the minimal estimated prediction error. There are two ways to estimate this prediction error: 
cross-validation which is computationally intensive, or approximation methods such as the $C_{p}$ type statistics.
When a log-likelihood loss function is used, the latter leads to the well-known Akaike information criterion (AIC). Another commonly used information based criterion  is Bayesian information criterion (BIC), which imposes a larger penalty than the AIC.  

\subsection{LASSO} \label{LASSO}

LASSO is an $L_{1}$ penalization method proposed by Tibshirani \citep{tibshirani1996regression} 
for building parsimonious models when the performance of classical methods such as stepwise regression or best subset selection is not satisfactory. For linear regression  LASSO solves a penalized least squares problem along the regularization path, where the regression coefficients associated with unimportant covariates shrink to exactly zero while granting non-zero coefficients for important covariates. 
The theoretical properties of LASSO have been extensively studied under the linear regression model. 
Meinshausen and B\"uhlmann \cite{meinshausen2006high} showed consistency of LASSO under the neighborhood stability condition, when the true non-zero coefficients are sufficiently large in absolute value. This condition is equivalent to the irrepresentable condition used by Zhao and Yu \cite{zhao2006model}. 
Although some of these theoretical conditions might be difficult to achieve in practice,  LASSO has gained numerous attention as a technique to reduce dimensionality and construct predictive models. One of the main reasons for its popularity is its computational simplicity, involving convex optimization only. 
 In the SEER-Medicare linked data,  the codes (covariates) are categorical rather than continuous,  alternative versions of LASSO have been proposed to handle grouped and categorical data. For example, Yuan and Lin \citep{yuan2006model} introduced group LASSO to include or exclude the grouped variable by replacing the $L_{1}$ penalty with $ \|{\bm{\beta}}\|_{K} = (\bm{\beta}^{T}\bm{K}\bm{\beta})^{1/2}$, where $\bm{K}$ is a symmetric positive definite matrix. In a more recent paper,  Gertheiss and Tutz \citep{gertheiss2010sparse} introduced a different penalty function $J(\bm{\beta}) = \sum_{i>j}w_{ij}|\beta_{i}-\beta_{j}|$, which is similar to the adaptive LASSO \citep{zou2006adaptive}. 
 Although we found  R codes   for group LASSO as well as adaptive LASSO in the survival context, the codes failed to work due to the scale of our data.  On the other hand, 
 the `glmnet' implementation of LASSO has been widely used and is able to handle large data sets.
 
Tibshirani \citep{tibshirani1997lasso} extended LASSO to the Cox regression model, where the log partial likelihood is penalized by $\lambda \|{\bm{\beta}}\|_1$. In fitting the PCSH model, the Cox regression software is used, and we apply the same LASSO algorithm as proposed in \citep{tibshirani1997lasso}. 
The penalty parameter $\lambda$ can be determined by different methods, and in the following we consider:
\begin{itemize}
\item CV10: $\lambda$ associated with the minimum 10-fold cross-validated (CV)  negative predictive log partial likelihood (referred to as `error' in the following);
\item CV+1SE: $\lambda$ associated with the minimum 10-fold CV error plus one standard error of the CV estimated errors; 
\item min AIC / BIC: $\lambda$ associated with the minimum AIC or BIC criteria; 
\item elbow AIC / BIC: $\lambda$ associated with the largest descent in AIC or BIC. 
\end{itemize} 
In the above under the Cox model, 
the AIC is defined as $-2\log(L) + 2s$, where  $L$ is the partial likelihood and $s=|S(\hat{\bm\beta})|$ is the number of non-zero regression coefficients, i.e.~the size of the active set $S(\hat{\bm\beta})$ \cite[]{verweij1993cross, xu:etal:09}. 
BIC under the Cox model is defined as $-2\log(L) + 2s\log(k)$, where $k$ is the number of observed uncensored events \cite[]{voli:raft}. 
We apply these definitions to the PCSH model, where $k$ would be the number of observed events from the cause of interest. 
The `elbow' criteria are described in Tibshirani {\it et al}.~\cite[]{tibsh:etal:gap} as a way to avoid over-selection in practice.  

We note that while Fu \textit{et al.}~\cite{fu:etal:2016} considered the LASSO and other penalized approaches under the PSDH model, we were not able to apply the associated R package `crrp'  to the linked SEER-Medicare data as it ran out of memory. 

\subsection{Boosting}\label{Boosting}

Freund and Schapire \cite{freund1995desicion} introduced the AdaBoost algorithm to solve classification problems by combining rough and moderately accurate `rules of thumb' repeatedly.  Later, Friedman \cite{friedman2001greedy} developed boosting methods for linear regression  as a numerical optimization method to minimize the squared error loss function. Boosting can be viewed as a gradient descent optimization algorithm in function space, and is essentially the same as the matching pursuit algorithm in signal processing \cite{mallat1993matching}. 
B\"uhlmann \cite{buehlmann2006boosting} proved that boosting with the squared error loss is consistent in high-dimensional linear models, where the number of predictors is allowed to grow as fast as exponential to the sample size. 

For the PSDH model with high dimensional data Binder \textit{et al}.~\citep{binder2009boosting} proposed a likelihood based boosting approach, where the likelihood is the same as the partial likelihood in \cite{fine1999proportional} for complete (i.e.~no censoring) data, but otherwise with weights in the risk sets to account for censoring:
\be\label{eq:fglik}
L(\beta) = \prod_{i=1}^n \left[ \frac{ \exp (\beta' Z_i) } { \sum_{l\in R_i} w_l(X_i) \exp (\beta' Z_l) } \right]^{ I(\delta_i \epsilon_i=1) }, 
\ee
where $R_i = \{l: X_l \geq X_i \ \mbox{or} \  \delta_l \epsilon_l >1 \} $ is the risk set consisting of individuals who have not had any event or who have had an event of other causes, and $w_l(t) = \hat{G}(t) I(t \geq X_l) \delta_l / \hat{G}(X_l) + I(t < X_l)$ (Binder \textit{et al}.~missed the second summand) where $\hat{G}$ is the Kaplan-Meier estimate of $P(C>t)$. 
The number of boosting steps $\gamma$, which is the main tuning parameter for this approach, can be  determined by the following criteria:
\begin{itemize}
\item CV10: $\gamma$ associated with the minimum  10-fold CV negative predictive log partial likelihood;
\item min AIC / BIC: $\gamma$ associated with the minimum AIC or BIC criteria; 
\item elbow AIC / BIC: $\gamma$ associated with the largest descent in AIC or BIC. 
\end{itemize} 
The definitions of the criteria are  similar to those under the PCSH model above, with the likelihood (\ref{eq:fglik}) returned by the `crr()' function in R package `cmprsk'.

\section{Simulations}\label{Simulations}

\subsection{Setup}

To investigate the performance of LASSO and boosting  under the  PCSH and PSDH models, respectively, we conducted comprehensive simulation studies with both continuous and dichotomized covariates in competing risks data. We assumed $J=2$, and we considered sample size $n=500$ and number of covariates $p=20$, 500, and 1000. We repeated each simulation setting $100$ times. 

For continuous covariates, 
the covariate vector for each subject 
  was generated for the following correlation structures: 
\begin{itemize}
\item[] 1)  Independent:  each covariate  was independently generated from $N(0, 1)$;
\item[] 2) Exchangeable: the covariate vector  was generated from a multivariate normal distribution with mean zero, marginal variance of one, 
and a block diagonal covariance 
matrix - each block of size $10$ and within a block the pairwise correlation $\rho(i, i^{\prime})=0.5$.
\item[] 3) AR(1):  the covariate vector was generated from a multivariate normal distribution with mean zero, marginal variance of one, 
and a block diagonal covariance 
matrix - each block of size $10$ and within a block the pairwise correlation $\rho(i, i^{\prime})=0.5^{|i-i^{\prime}|}$.
\end{itemize}
For binary covariates, 
the covariate vector  
was first generated the same as in the above, then dichotomized at threshold $a$, with $<a$ coded as 1 and 0 otherwise.
We considered $a=0$, to give a balanced binary distribution, and $a=-1$, to give a relatively sparse 16\% of 1's.
We set the number of non-zero regression coefficients, i.e.~the size of the active set, 
to be $s_1=5$ and $s_2=3$ for causes 1 and 2, respectively. We let $\bm{\beta}_{1, 1\cdots, 5} = (1.96,-0.79, -0.5, -1.35,  1.29)$, $\bm{\beta}_{2, 11\cdots, 13} = (-1.16,-0.86, 0.5)$ and the rest of the $\bm{\beta_1}$ and $\bm{\beta_2}$ values were zero. 
These $\bm{\beta}$ values were used under both the PCSH and the PSDH models.

To simulate survival outcomes under the PCSH model we followed the approach described in \citep{beyersmann2009simulating}; that is, we simulated the event time $T$ first, then we simulated the cause $\epsilon$ given $T$.  We assumed the baseline hazard functions for type 1 and 2 failures to be $\lambda_{01}(t) = 0.15$  and $\lambda_{02}(t) = 0.10$, respectively. The overall (not cause-specific) cumulative hazard function for $T$ was then
$ 
\Lambda(t | \bm{z}) 
= t \{ \lambda_{01}\exp(\bm{\beta_{1}}' \bm{z})  + \lambda_{02}\exp(\bm{\beta_{2}}' \bm{z}) \}  
$, and $T$ was generated using the fact that 
     $    U = \exp(-\Lambda(T)) \sim  U(0,1)$ given $ \bm{z}$. 
  The cause $\epsilon$ was generated proportional to the cause-specific hazard function, i.e.~
     $P(\epsilon =1 | \bm{z}) = \lambda_{01}\exp(\bm{\beta_{1}}' \bm{z}) / \{ \lambda_{01}\exp(\bm{\beta_{1}}' \bm{z})  + \lambda_{02}\exp(\bm{\beta_{2}}' \bm{z}) \}  $. 
Under this  model, the true CIF for cause $j$ was  
\begin{eqnarray}\label{trueCIF}
\mbox{CIF}_{j}(t | \bm{z} ) = \int_{0}^{t} S(u | \bm{z})\lambda_{0j}\exp( \bm{\beta_j}' \bm{z} )du 
= \lambda_{0j}\exp( \bm{\beta_j}' \bm{z} )   \frac{e^{tM}}{M},
\end{eqnarray} 
where $M= -\{\lambda_{01}\exp( \bm{\beta_1}' \bm{z} )  + \lambda_{02}\exp( \bm{\beta_2}' \bm{z} ) \}$. 
The censoring times were generated from $U(0, 20)$, which resulted in an average event rate of $45.8\%$ for cause 1 and $33.6\%$ for cause 2 with continuous covariates,
an average event rate of $51.8 \%$ for cause 1 and $27.2 \%$ for cause 2 with balanced binary covariates, 
and an average event rate of $59.8\%$ for cause 1 and $17.8\%$ for cause 2 with sparse binary covariates.. 
 
To simulate  under the PSDH model we followed the approach described in  \citep{fine1999proportional}. 
The CIF for failure from cause 1 was given by
\begin{eqnarray} \label{subdist1}
\mbox{CIF}_1 (t | \bm{z} ) = P (T \le t, \epsilon =1 | \bm{z} ) = 1 - \{1 - p (1- e^{-t} ) \}  ^{\exp(\bm{\beta_1' z})},  
\end{eqnarray}  
where we used $p=0.6$.
As this was a subdistribution function, with a point mass $1- \mbox{CIF}_1(\infty | \bm{z} )$ at infinity, the proper distribution function that was used to generate $T$ was 
$F(t | \bm{z} ) = \mbox{CIF}_1(t | \bm{z} ) / \mbox{CIF}_1(\infty | \bm{z} )$, so that $F(T) \sim  U(0,1)$ given $ \bm{z}$. 
Note that $P( \epsilon =1 | \bm{z} ) = \mbox{CIF}_1(\infty | \bm{z} ) $, and $P( \epsilon =2 | \bm{z} ) = 1- P( \epsilon =1 | \bm{z} )$. 
Finally the event times for failure from cause 2
were generated according to an exponential distribution with rate $\exp(\bm{\beta_{2}' z})$. 
The censoring times were generated from $U(0, 20)$, resulting in   
 an average event rate of $53.5\%$ for cause 1 and $35.1\%$ for cause 2 with continuous covariates, an average event rate of $55.8\%$ for cause 1 and $33.4\%$ for cause 2 with balanced binary covariates, 
 and an average event rate of $55.5\%$ for cause 1 and $33.5\%$ for cause 2 with sparse binary covariates. 

\subsection{Results}

\begin{figure}[!htb]
  \caption{The (smoothed) empirical distribution of $\widehat{\cif}_1(2)$, estimated under the PCSH model with LASSO, for continuous covariates. The three columns correspond to $p = 20$, 500, and 1000. The rows correspond to different ways of selecting $\lambda$, from top to bottom: 1) CV10, 2) CV+1SE, 3) minimum AIC, 4) minimum BIC, 5) elbow AIC and 6) elbow BIC. The true $\cif_1(2 | z_{0})=0.32$. } 
\label{CSH_LASSO_CIF1_2yr}
  \centering
    \includegraphics[width=12cm]{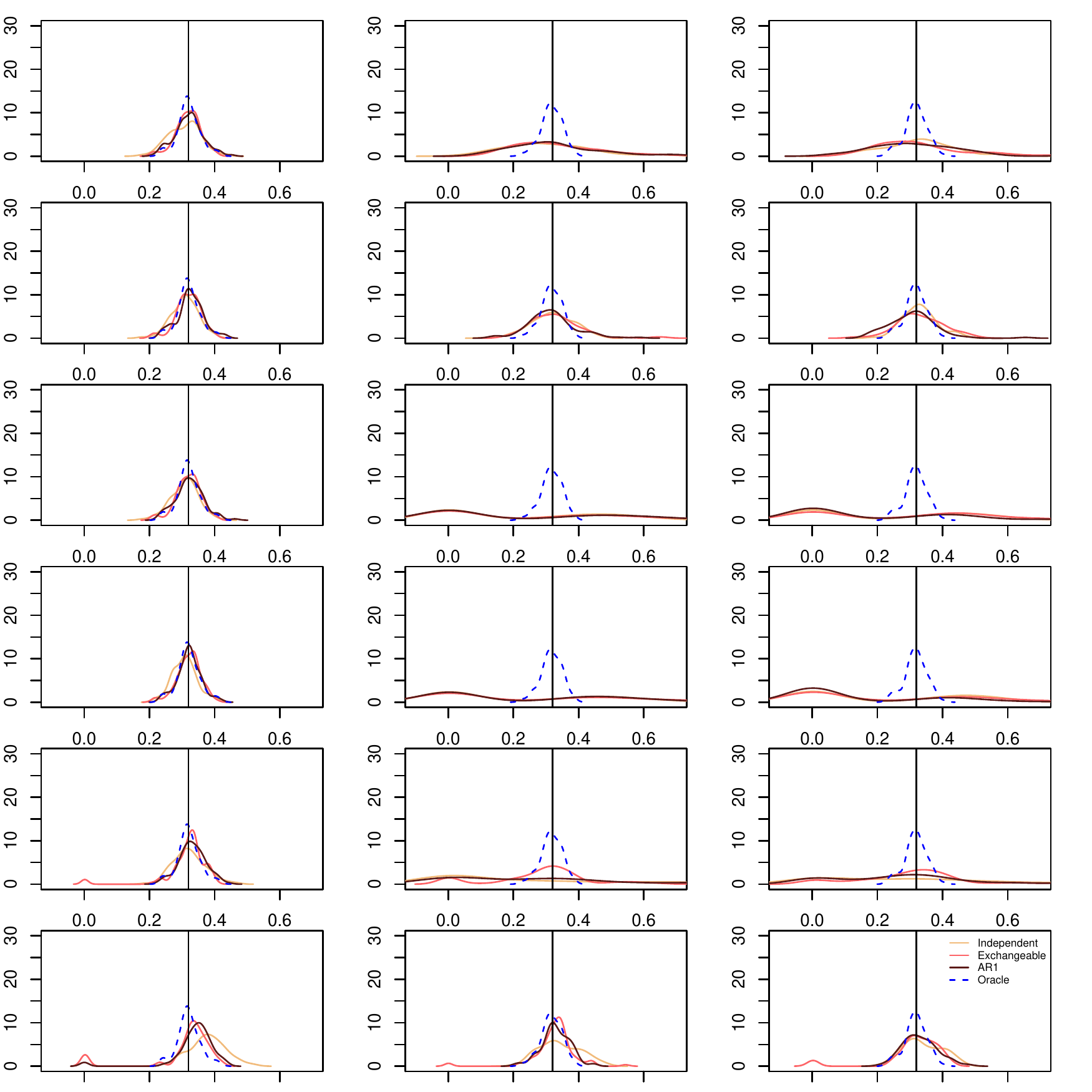}
         \end{figure}

 We evaluate the performance of prediction at a given covariate vector value $\bf{z}_{0}$. We set $\bm{z_{0}}=(0.5, \cdots, 0.5)_{1\times p}$ for the continuous case; and  for all the binary cases each element of $\bf{z}_{0}$ was independently drawn with a fixed seed from Bernoulli distribution with $p=0.5$. 
Figures \ref{CSH_LASSO_CIF1_2yr} - \ref{FG_boosting_CIF1_2yr_sparse_binary} show the empirical distributions of the estimated $\cif_1(2)$ over the 100 simulation runs, where the vertical line marks the true $\cif_1(2)$; the empirical distributions were plotted using the R function  `density()'. 
The PCSH model with LASSO was used to estimate $\cif_1(2)$ in Figures \ref{CSH_LASSO_CIF1_2yr} - \ref{CSH_LASSO_CIF1_2yr_sparse_binary}, and the PSDH model with boosting was used in Figures \ref{FG_boosting_CIF1_2yr} - \ref{FG_boosting_CIF1_2yr_sparse_binary}.

In the figures the blue dashed lines are for the oracle estimator, which fits the exact true active set $S(\beta)$. The oracle estimator varied extremely slightly when the three correlation structures for $Z$ were generated separately, which appeared to be due to Monte Carlo variation, and the one under the AR(1) structure is plotted here. It is seen that the distribution of the oracle estimator is more concentrated for the balanced binary covariates than for the sparse binary covariates, which reflects the `effective sample size' that is reduced with the sparse binary covariates.
The solid lines are the estimated $\cif_1(2)$ under each model after regularization using LASSO or boosting, with different colors representing different correlation structures of $Z$. 

Under the PSDH model using LASSO to regularize, the performances were generally not satisfactory as compared to the oracle estimator. The worst performances were seen when using minimum AIC and BIC to choose the penalty parameter; some of these results were so extreme that `density()' failed to work and these were instead shown in the Supplemental Materials using boxplots.  Elbow BIC appeared to perform the best for continuous covariates, but not so for  binary covariates  even when $p=20$. CV10 had the best performance for binary covariates for  $p=20$, but it too deteriorated  for $p=500$ and 1000. 

Under the PSDH model using boosting, in Figure \ref{FG_boosting_CIF1_2yr} we see that for continuous covariates, the estimators performed reasonably well  when CV10 or minimum AIC/BIC was used to choose the number of boosting steps; with CV10 the estimation was perhaps the best. The performance deteriorated with (balanced) binary covariates for $p=500$ and 1000. For the sparse binary covariates, even the oracle estimator had a very wide spread, with the performance of the regularized estimator under the independent structure  the worst of all. 
We note that in B{\"u}hlmann \cite{buehlmann2006boosting} simulation studies (Table 1) the mean squared error for boosting with correlated design was also smaller than that with uncorrelated design, and their Figure 1 showed that boosting tended to select more covariates in the uncorrelated design than the correlated design.

The Supplemental Materials provide the results of variable selection; although this was not the main goal for our application and it was difficult to achieve good model consistency (i.e.~selection), they help to explain the prediction accuracy. When the selection is extremely poor, for example a couple of hundred false positives, then the prediction results were very poor as well. Boosting had no more than five false positives in all cases.


\section{SEER-Medicare linked data}\label{realdata}
 
 We randomly split the SEER-Medicare dataset into approximate equal-sized training ($n=28,505$) and test ($n=28,506$) datasets. 
As a first step we excluded binary claim codes having less than 10 ones, with rest all being zero. 
We then used univariate screening  to further reduce dimensionality, eliminate noise and increase the performance of subsequent variable selection methods \citep{wasserman2009high}.

\subsection{PCSH model with LASSO}

Univariate screening under the  PCSH model with $p$-valued cutoff of 0.05  gave $p_1=2188$ and $p_2=1079$ claim codes for non-cancer and cancer mortality, respectively. 
For each type of mortality, we applied LASSO under the PCSH model described earlier  on the training data with the above  pre-screened claim codes plus the clinical and demographic variables. 
Based on the simulation results, CV10 was used to choose the penalty parameter. 
 The final  model contained 143  predictors for non-cancer mortality, and 9  predictors for cancer mortality.  
Since the regression coefficients  from LASSO are biased, we refit the PCSH model with the  selected predictors.  
 
In order to evaluate the resulting prediction model on the test data, we first calculated the risk score $\hat\beta'_j Z$ for each patient in the test data, $j=1, 2$. For each mortality type $j$, 
we then divided the test set into 4 risk strata: low (L), median low (ML), median high (MH) and high (H) according to the quartiles. Combining the two types of mortalities, we formed a total of 16 strata for their predicted CIF.  
Using the average $Z$ values in each of the 16 strata, we plotted their predicted CIF's for both cancer and non-cancer mortality (see Supplemental Materials). 
It was clear that instead of 16 groups, 5 distinct risk groups emerge for both cancer and non-cancer mortalities. 
We note that these are not the same 5 groups for the two types of mortality, and Table \ref{reclassification} provides the definition for each of them. It is perhaps not surprising to see that each mortality risk was most influenced by the corresponding cause-specific risk, and also secondarily by its competing risk.

In Figure \ref{CIF_overlap1} we plot the nonparametric CIF (solid lines) and its 95\% confidence intervals  (shaded) for each of the above 5 risk groups for each type of mortality. 
 The  clear separation of the 5 groups show the usefulness of the final PCSH model in classifying patients according to their different prognosis for both cancer and non-cancer mortalities. 
 For comparison purposes we also plot the predicted CIF for each of the 5 risk groups. While the prediction is more accurate for the non-cancer CIF, the predicted cancer CIF seems less accurate especially for the high (H) risk group.

\subsection{PSDH model with boosting} 
Univariate screening with $p$-value  cutoff of 0.05 under the  PSDH model initially gave $p_1= 4634$ and $p_2=6088$ claim codes for non-cancer and cancer mortality, respectively. We further reduced the dimension by retaining only the top 2000 claim codes (ranked by $p$-value) for each type of mortality, to be comparable with the fitting of the PCSH model above, as well as for the boosting algorithm to be able to run on our Dell R630 computer (two Intel Xeon E5-2660 v3 2.6GHz, each processor with 10 cores (20 threads) for a total of 20 cores (40 threads), 
128GB of DDR3). We applied boosting under the PSDH model to the training data with these claim codes plus the clinical and demographic variables. Although CV10 performed best in our simulation results, AIC was a close second especially for binary covariates and was less computationally intensive for this procedure where boosting itself was computationally intensive already. Therefore AIC was used to choose the optimal step. The final model contains $53$ predictors for non-cancer mortality, and $13$ predictors for cancer mortality. We refit the PSDH model with the selected predictors to obtain the unbiased estimator. 

Similar to the abve, we  calculated the risk score $\hat\beta'_j Z$ for each patient in the test data, $j=1, 2$. For each mortality type $j$, we divided the test set into 5 risk strata: low (L), median low (ML), median (M), median high (MH) and high (H) according to the quartiles. Again the classification was not the same 5 groups for the two types of morality.  In Figure \ref{CIF_overlap2}, we plot the nonparametric CIF (solid lines) and its 95\% confidence intervals  (shaded) for the above 5 risk groups for both non-cancer and cancer mortality.   For comparison purposes we also plot the predicted CIF for each of the 5 risk groups. 
The prediction was still more accurate for the non-cancer CIF, and less so for cancer CIF especially for the high (H) risk group; however, compared to the PCSH model the separation of the 5 groups was less even, making the results perhaps less desirable for clinical use.

\section{Discussion} \label{Discussion}

The rapid accumulation of data across many fields, medicine in particular, has created unique challenges in statistics. The distinct issues with  high dimensional data have come to be recognized recently, 
 including for example, 
the rapid noise accumulation,  the unrealistic independence assumption, 
and the necessity for novel robust data analysis methods \cite[]{2014challenges}.
While researchers work to meet these challenges, some of the methods proposed in the literature do not necessarily scale well to large data sets. 
In this paper, we considered the feasible implementations of  statistical learning methods under the PCSH and PSDH models. We empirically studied their performance in variable selection and prediction  through comprehensive simulations in both low- and high-dimensional settings with different covariate structures.  By applying the methods to analyze a rich  dataset with claim codes describing disease diagnoses, surgical procedures, hospitalization and outpatient activities, we created an individualized patient prediction tool aimed at helping prostate cancer patients and their physicians to better understand the prognosis for both cancer and other morbidities, which can in turn aid in clinical decision making. 

For the linked SEER-Medicare data that we have considered, the PCSH model appears to outperform the PSDH model in terms of classifying patients into different risk groups for both cancer and non-cancer mortality. PCSH outperforming PSDH  has not been previously observed; in the limited comparisons that we are aware of in the literature, the two seem to give somewhat comparable results \cite[]{geskus}. While the PSDH model was proposed in order to associate the CIF due to one cause directly with the covariates without having to specifically model the other causes, the PCSH model might be more flexible precisely due to the fact that it allows different modeling of different causes in the denominator of the CIF. This is certainly worth future investigation. 
We also note that while the proportional hazards assumption is used in both models, there has been recent work considering other modeling approaches such as the additive hazards in the presence of competing risks \cite[]{zheng:iv}.  

Finally, we note that in the high-dimensional context  methods developed for continuous data may behave differently for binary data especially if sparsity presents. 
In a recent paper Mukherjee \textit{et al.}~\citep{mukherjee2015hypothesis} showed that when a binary design matrix is sufficiently sparse,
 no signal can be detected irrespective of its strength. 
This finding echos the challenges that we have observed in our simulation studies.

\vskip .3in
\centerline{\bf\large Acknowledgement}

\vskip .15in
This research was supported by a grant from the American Society of Clinical Oncology (ASCO), and was partially supported by the National Institutes of Health Clinical and Translational Science Award (CTSA) UL1TR001442.

\bibliographystyle{plain}
\bibliography{8248_stat_paper}

\clearpage

\begin{table}[ht]
\caption{Five risk groups ($n$ and \# events from test data) derived from the original 16 combinations of 4  non-cancer and 4 cancer cause-specific strata.}
\label{reclassification}
$$\begin{tabular}{cccll}
  \hline\hline
\multicolumn{1}{c}{Risk Groups}  &  \multicolumn{1}{c}{$n$}  &\multicolumn{1}{c}{\# events}  &\multicolumn{2}{c}{Cause-Specific} \\
  \hline
 Non-cancer mortality:  &&& Non-cancer: & Cancer: \\
 L & 7127 & 131 & L & L, ML, MH, H \\
ML & 7126 & 280 & ML & L, ML, MH, H \\
M & 7126 & 558 & MH & L, ML, MH, H \\
MH & 3117 & 594 & H & L, ML, MH\\
H & 4010 & 1022 & H & H \\
\hline
Cancer mortality:  &&& Non-cancer: & Cancer: \\
 L & 7135 & 40 & L, ML, MH, H  & L \\
 ML & 7118 & 51 & L, ML, MH, H  & ML \\
 M  & 7128 & 101 & L, ML, MH, H  & MH \\
 MH & 3115 & 134 & L, ML, MH & H \\
H & 4010 & 294 & H & H \\
\hline
 \end{tabular}$$
 \end{table}

\clearpage

 \begin{figure}[!htb]
  \caption{The (smoothed) empirical distribution of $\widehat{\cif}_1(2)$, estimated under the PCSH model with LASSO, for balanced binary covariates. The three columns correspond to $p = 20$, 500, and 1000. The rows correspond to different ways of selecting $\lambda$, from top to bottom: 1) CV10, 2) CV+1SE, 3) minimum AIC, 4) minimum BIC, 5) elbow AIC and 6) elbow BIC. The true $\cif_1(2 | z_{0})=0.11$. 
} 
\label{CSH_LASSO_CIF1_2yr_binary}
  \centering
    \includegraphics[width=12cm]{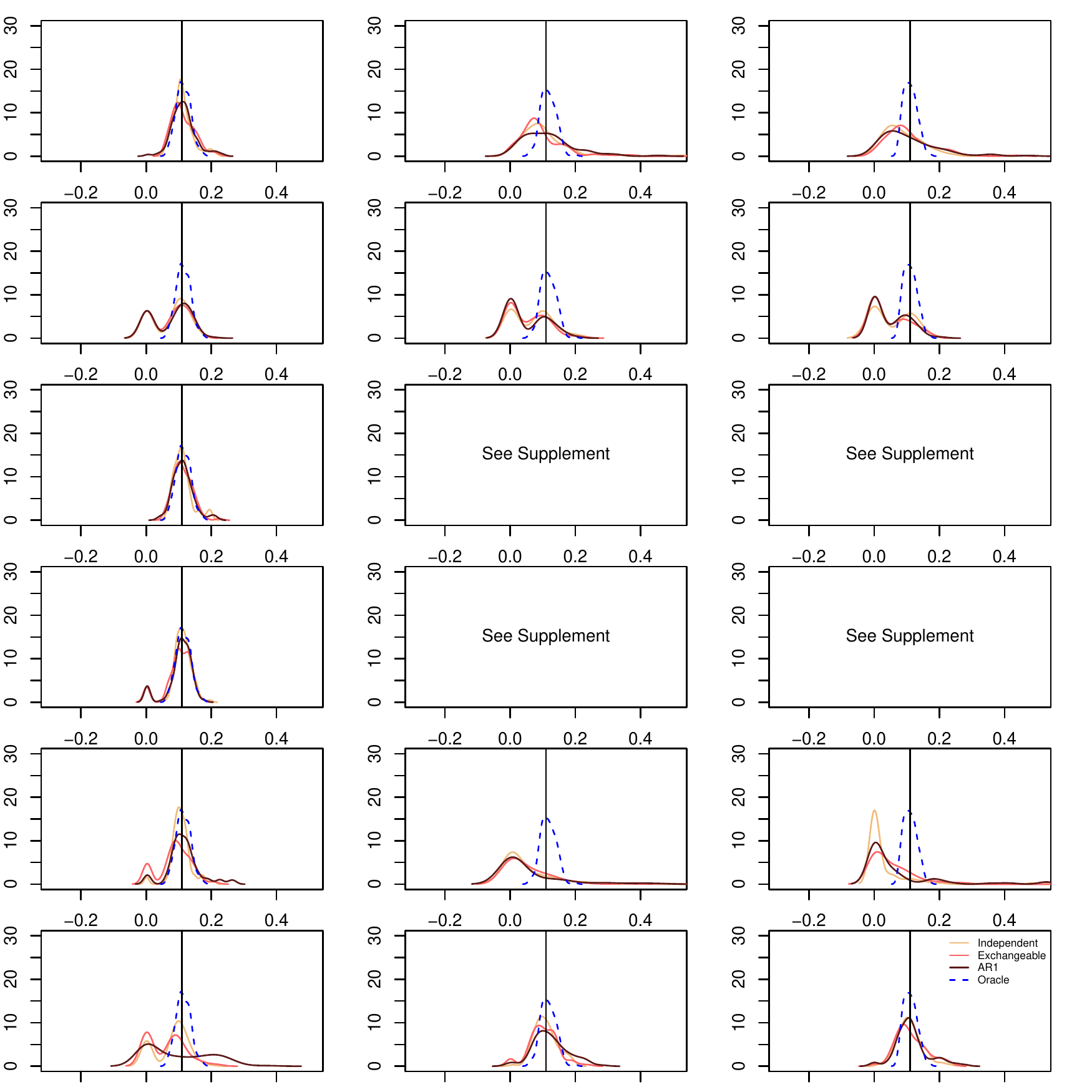}
         \end{figure}
   
 \begin{figure}[!htb]
  \caption{The (smoothed) empirical distribution of $\widehat{\cif}_1(2)$, estimated under the PCSH model with LASSO, for sparse binary covariates. The three columns correspond to $p = 20$, 500, and 1000. The rows correspond to different ways of selecting $\lambda$, from top to bottom: 1) CV10, 2) CV+1SE, 3) minimum AIC, 4) minimum BIC, 5) elbow AIC and 6) elbow BIC. The true $\cif_1(2 | z_{0})=0.11$. 
} 
\label{CSH_LASSO_CIF1_2yr_sparse_binary}
  \centering
    \includegraphics[width=12cm]{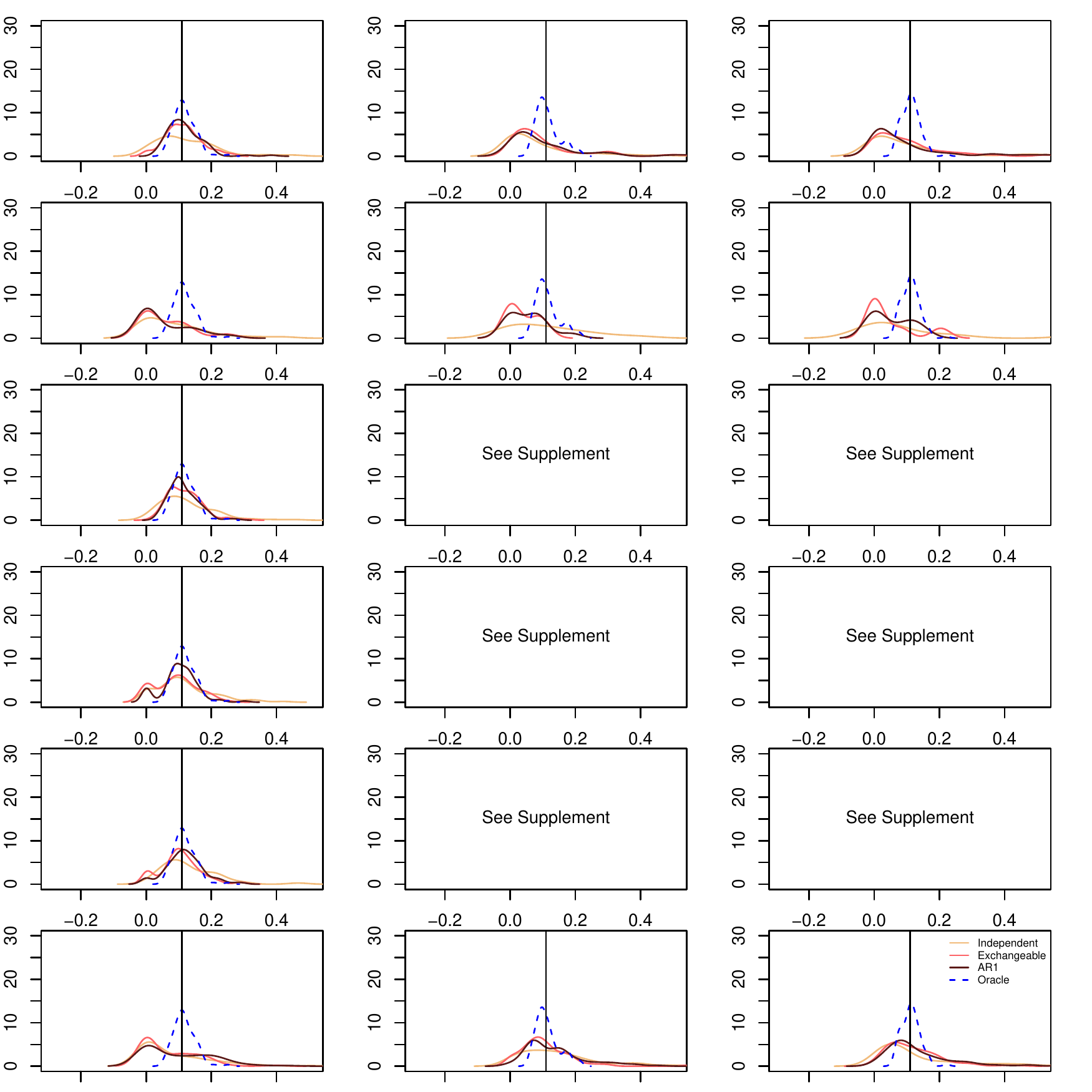}
         \end{figure}

 \begin{figure}[!htb]
 \caption{The (smoothed) empirical distribution of $\widehat{\cif}_1(2)$, estimated under the PSDH model with boosting, for continuous covariates. The three columns correspond to $p = 20$, 500, and 1000. The rows correspond to different ways of selecting $\gamma$, from top to bottom: 1) CV10, 2) minimum AIC, 3) minimum BIC, 4) elbow AIC and 5) elbow BIC. The true $\cif_1(2 | z_{0})=0.63$. 
}
\label{FG_boosting_CIF1_2yr}
  \centering
\includegraphics[width=12cm]{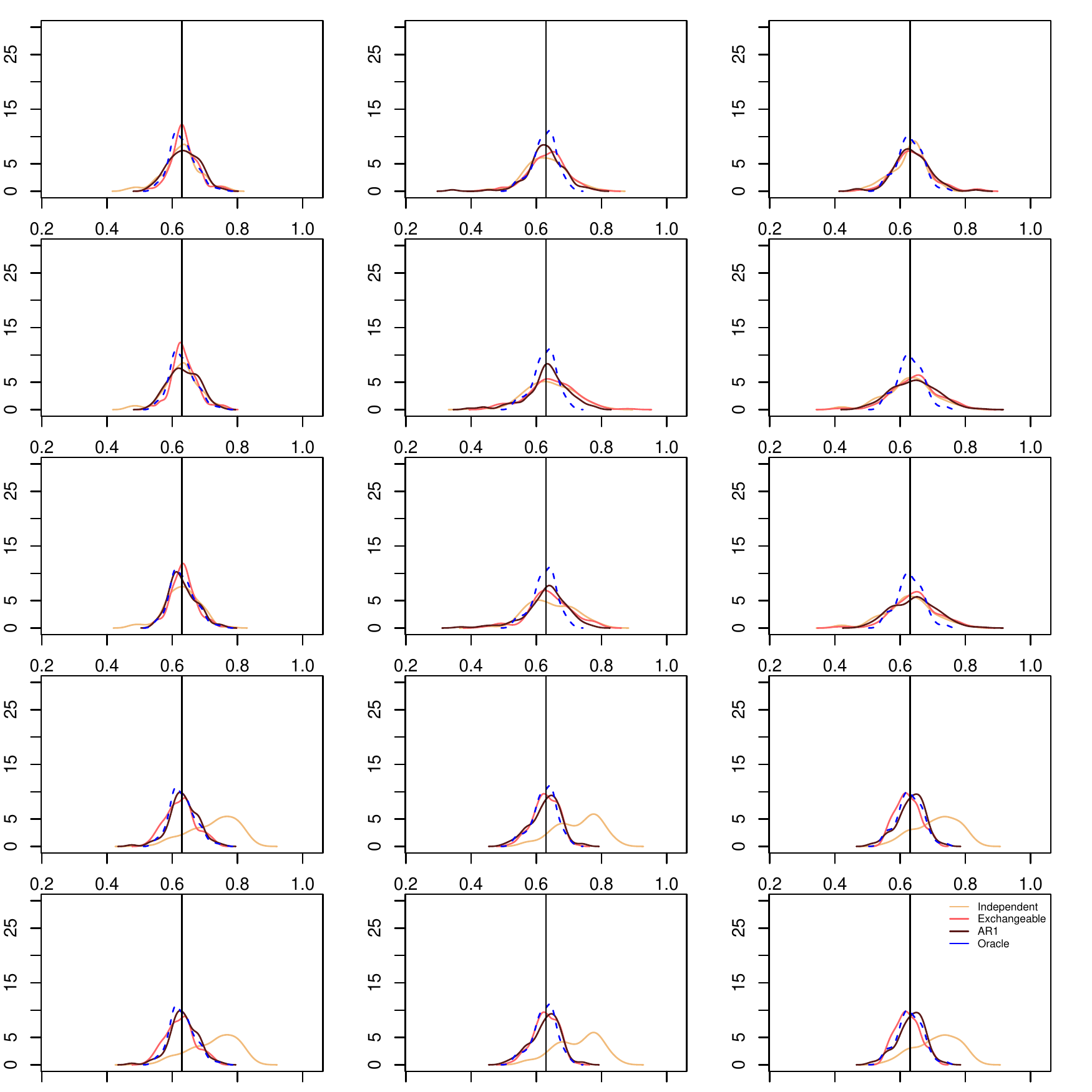}
    \end{figure}

  \begin{figure}[!htb]
 \caption{The (smoothed) empirical distribution of $\widehat{\cif}_1(2)$, estimated under the PSDH model with boosting, for balanced binary covariates. The three columns correspond to $p = 20$, 500, and 1000. The rows correspond to different ways of selecting $\gamma$, from top to bottom: 1) CV10, 2) minimum AIC, 3) minimum BIC, 4) elbow AIC and 5) elbow BIC. The true $\cif_1(2 | z_{0})=0.27$. 
}
\label{FG_boosting_CIF1_2yr_binary}
  \centering
\includegraphics[width=12cm]{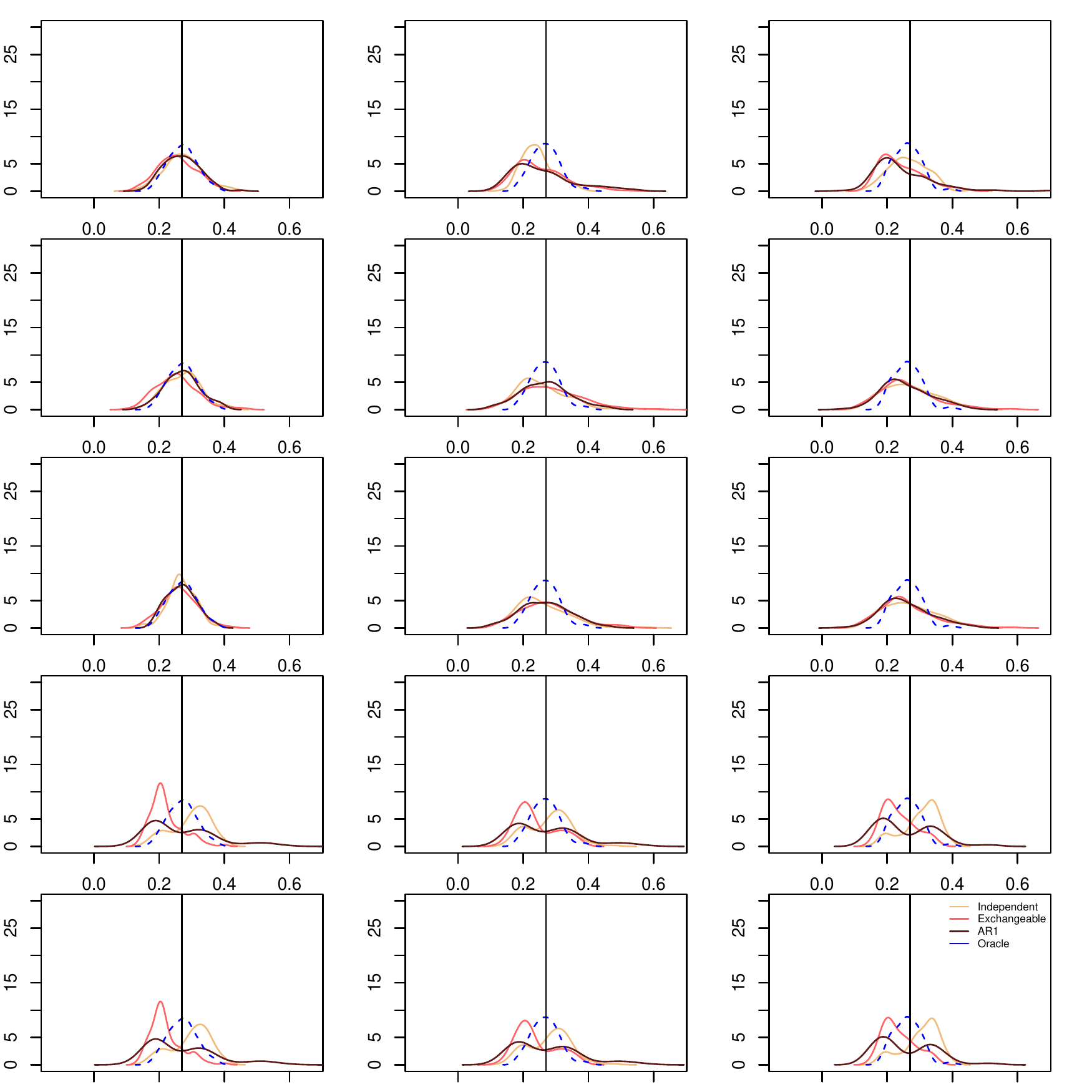}
    \end{figure}
    
  \begin{figure}[!htb]
 \caption{The (smoothed) empirical distribution of $\widehat{\cif}_1(2)$, estimated under the PSDH model with boosting, for sparse binary covariates. The three columns correspond to $p = 20$, 500, and 1000. The rows correspond to different ways of selecting $\gamma$, from top to bottom: 1) CV10, 2) minimum AIC, 3) minimum BIC, 4) elbow AIC and 5) elbow BIC. The true $\cif_1(2 | z_{0})=0.27$. }
\label{FG_boosting_CIF1_2yr_sparse_binary}
  \centering
\includegraphics[width=12cm]{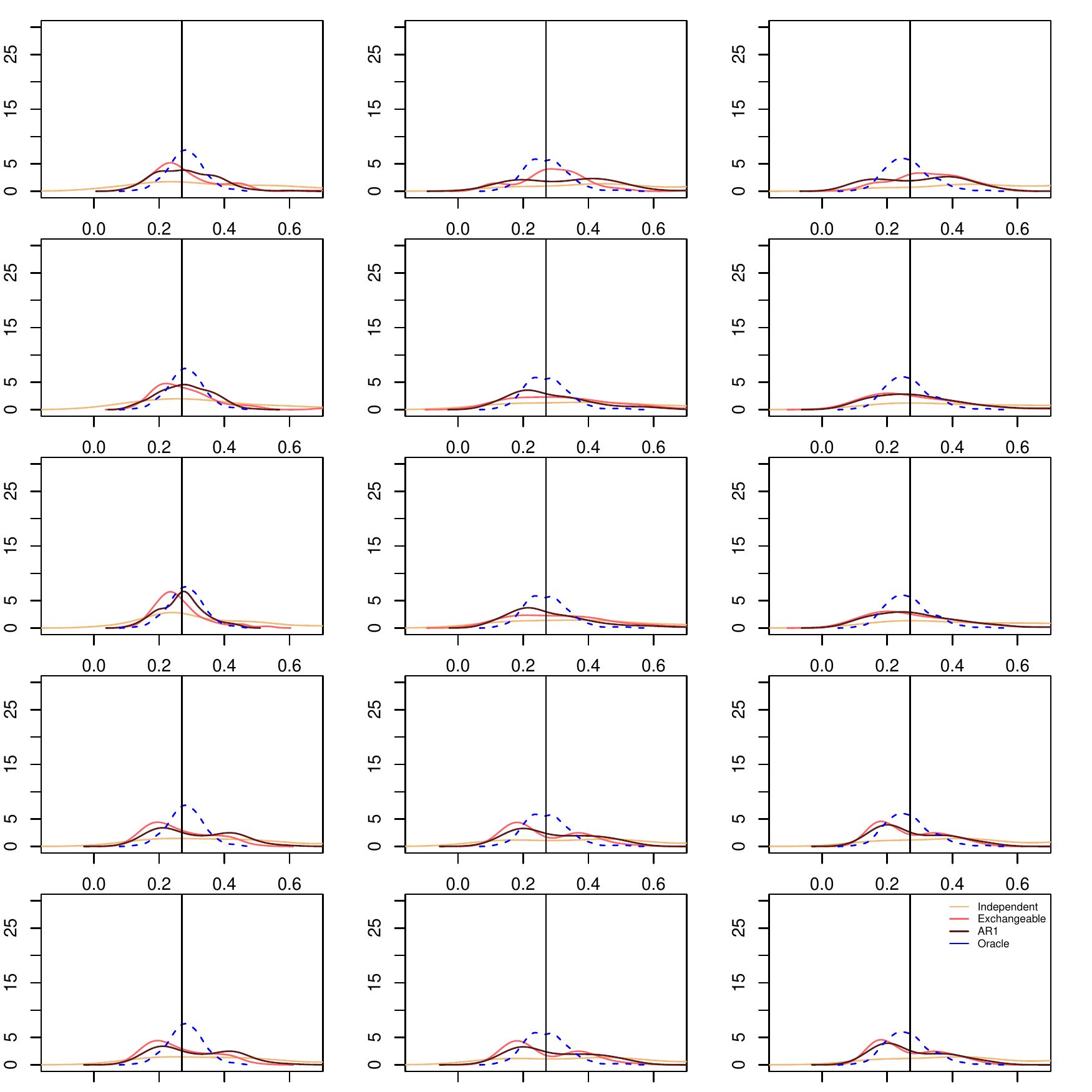}
    \end{figure}


\begin{figure}[!htb]
  \caption{Cumulative incidence functions for non-cancer (top) and cancer (bottom) mortalities, with classification and prediction based on the PCSH model. The shaded area is the 95\% pointwise confidence intervals based on the nonparametric estimate. 
  }
  \label{CIF_overlap1}
    \centering
      \includegraphics[width=8cm]{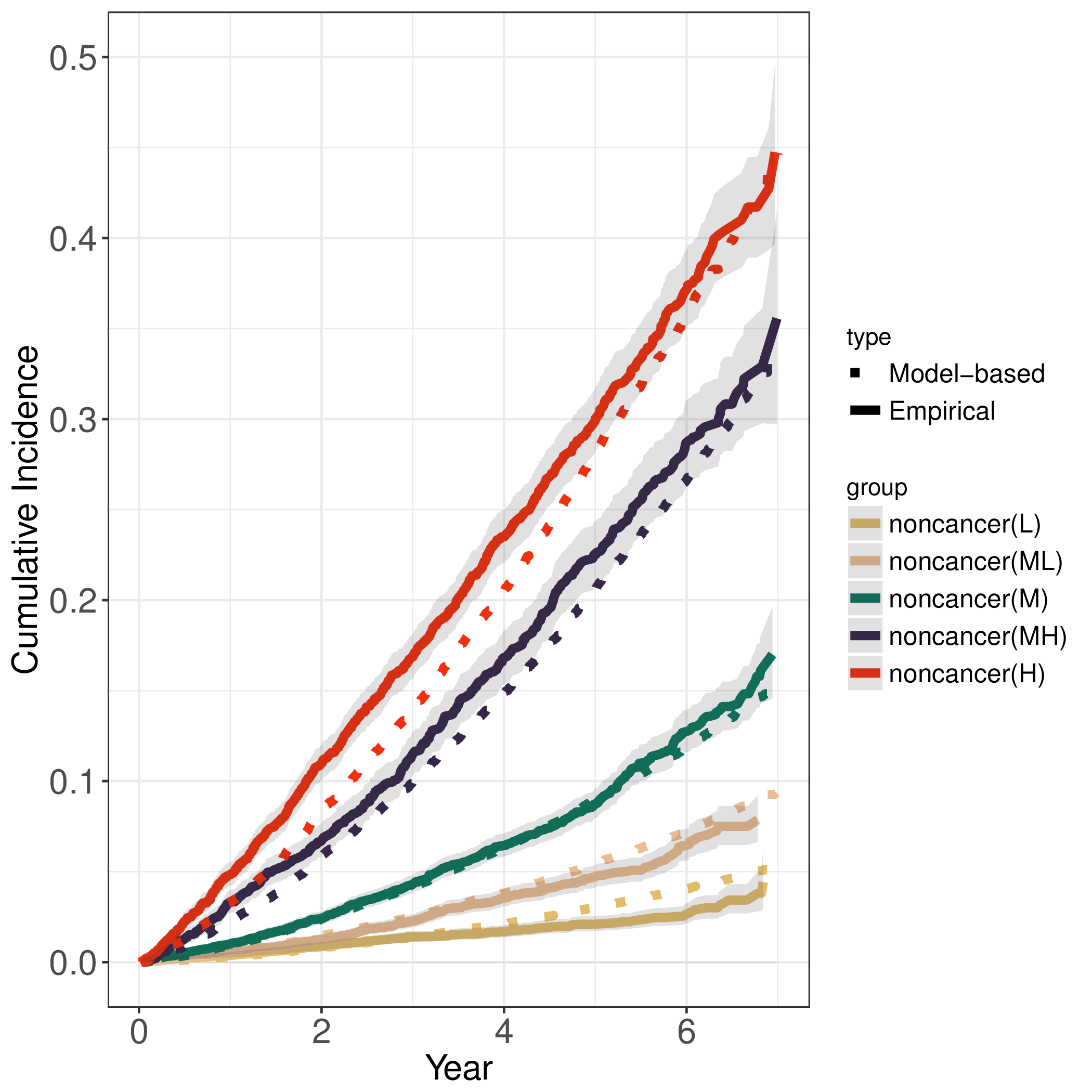}
      \includegraphics[width=8cm]{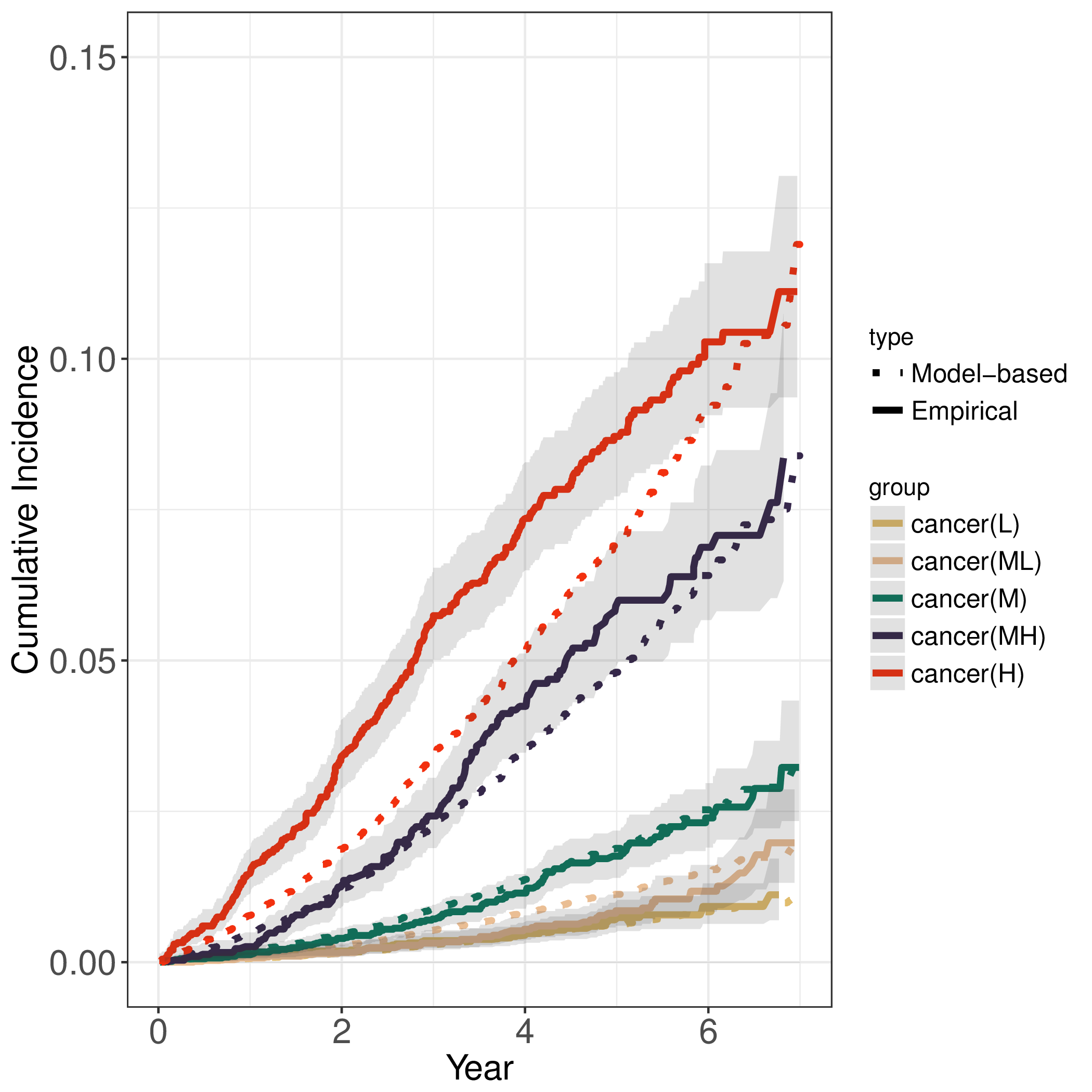}
   \end{figure}

\begin{figure}[!htb]
  \caption{Cumulative incidence functions for non-cancer (top) and cancer (bottom) mortalities, with classification and prediction based on the PSDH model. The shaded area is the 95\% pointwise confidence intervals based on the nonparametric estimate. 
 }
  \label{CIF_overlap2}
    \centering
       \includegraphics[width=8cm]{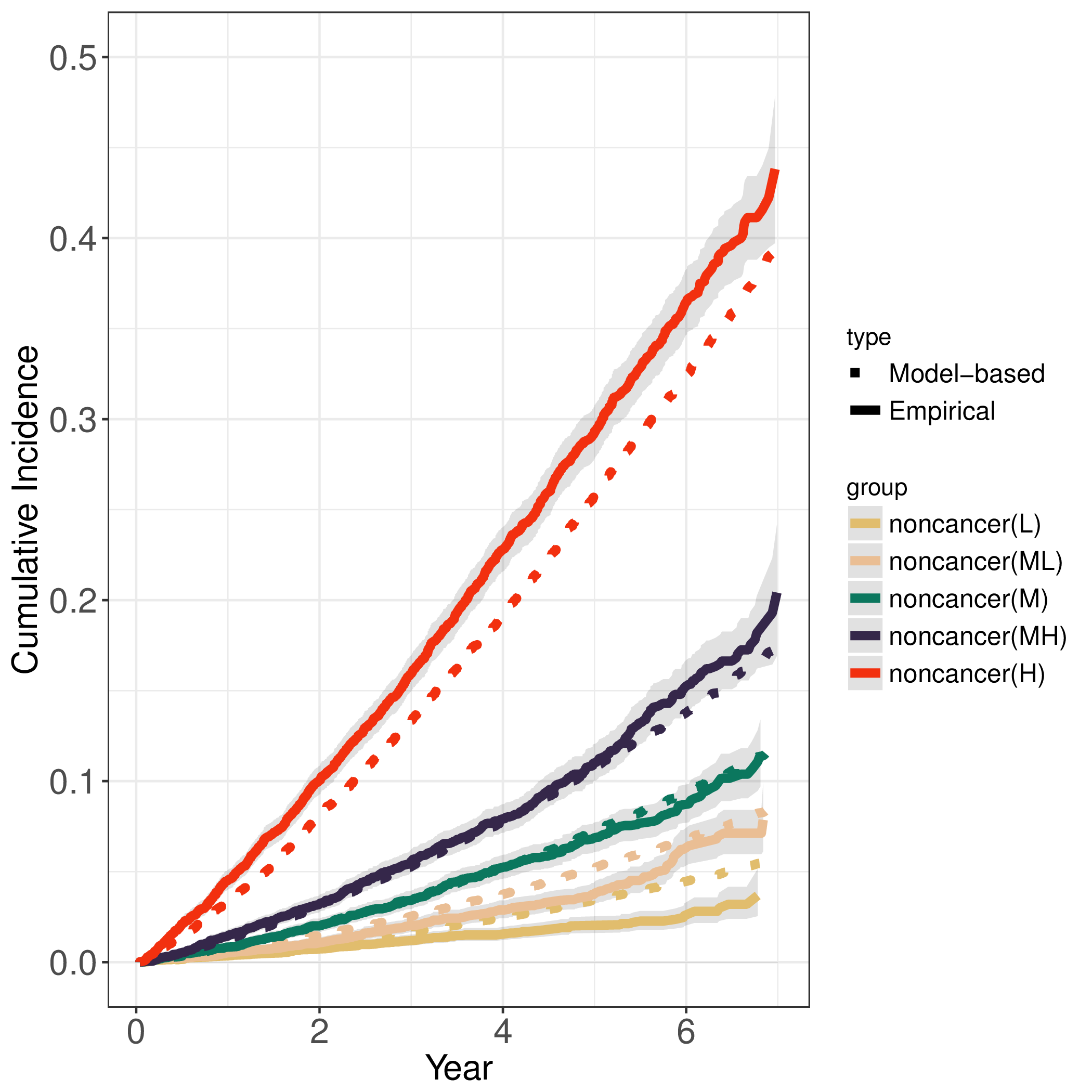}
      \includegraphics[width=8cm]{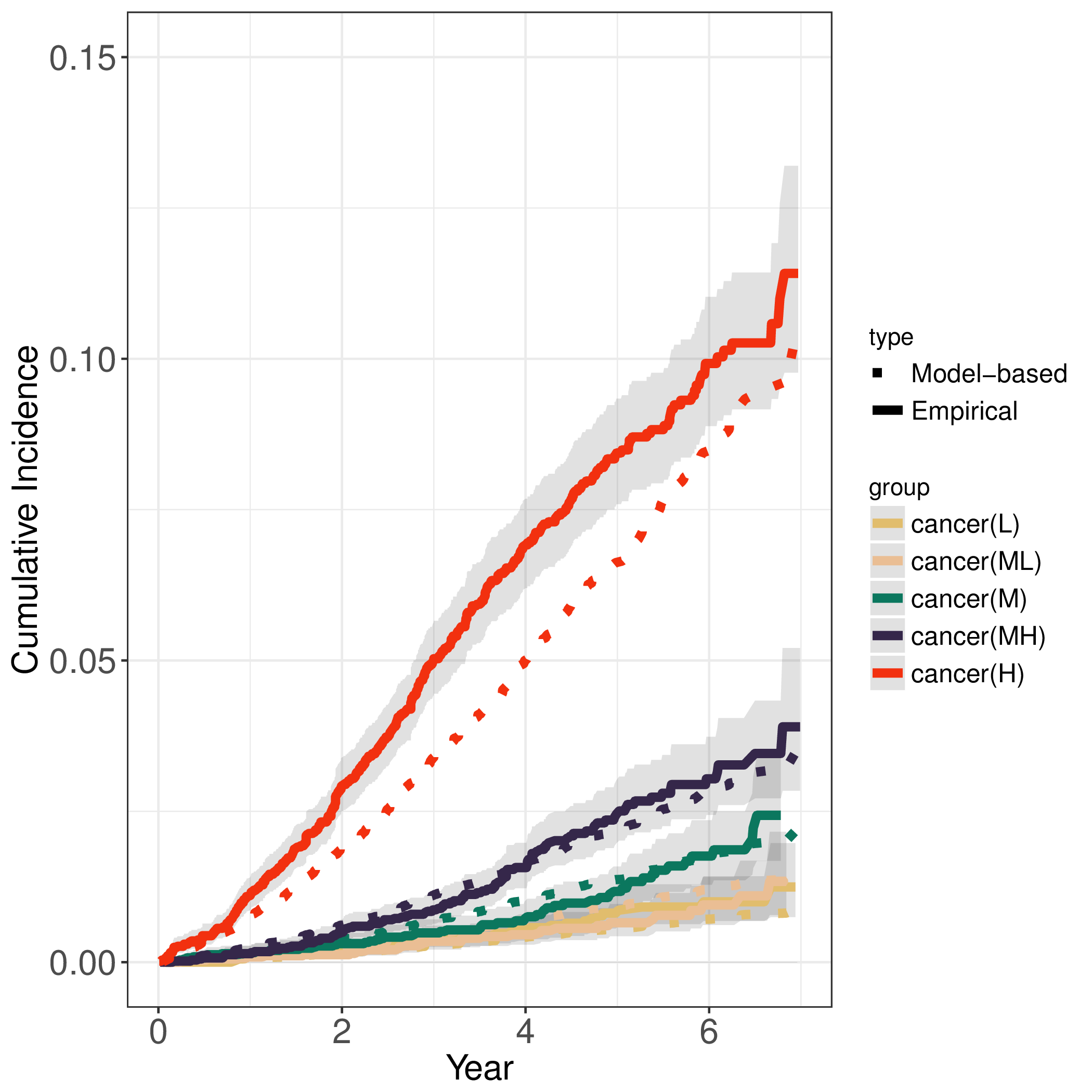}
  \end{figure}

\clearpage
\section{Supplemental Materials}

In the following tables, $|S(\hat{\beta})|$ is the size of the estimated active set, i.e.~number of non-zero estimated regression coefficients. The median number of selected variables are reported, and in () are the median absolute deviation (MAD) of selection.


\clearpage
\begin{table}[!htb]
\centering
\caption{The median (MAD) number of selected variables by LASSO under the PCSH model. 
The penalty parameter is chosen using  CV10.}
\label{CSH_lasso_selection1}
\small
\begin{tabular}{llll}
  \hline\hline
Cause 1 Continuous  &$|S(\hat{\beta})|$  & $\#$True Positives & $\#$False Positives \\ 
  \hline
  p=20 &&\\
  Independence & 12(1) & 5(0) & 7(1) \\ 
  Exchangeable & 12(2) & 5(0) & 7(2) \\ 
  AR1 & 12(1) & 5(0) & 7(1) \\ 
  \hline
    p=500 &&\\
 Independence & 35(7) & 5(0) & 30(7) \\ 
  Exchangeable & 35(5) & 5(0) & 30(5) \\ 
  AR1 & 37(6) & 5(0) & 32(6) \\ 
    \hline
    p=1000 &&\\
Independence & 41(8) & 5(0) & 36(8) \\ 
  Exchangeable & 44(7) & 5(0) & 39(7) \\ 
  AR1 & 41(7.5) & 5(0) & 36(7.5) \\ 
   \hline\hline
  Cause 1 Binary (balanced) &$|S(\hat{\beta})|$  & $\#$True Positives & $\#$False Positives \\ 
  \hline
  p=20 &&\\
  Independence & 11(2) & 5(0) & 6(2) \\ 
  Exchangeable & 12(2) & 5(0) & 7(2) \\ 
  AR1 & 12(2) & 5(0) & 7(2) \\ 
  \hline
  p=500 &&\\
  Independence & 22(6) & 5(0) & 17(6) \\ 
  Exchangeable & 27(6) & 5(0) & 22(6) \\ 
  AR1 & 25(6) & 5(0) & 20(6) \\ 
  \hline
  p=1000 &&\\
  Independence & 23(6) & 5(0) & 18(6) \\ 
  Exchangeable & 29(5) & 5(0) & 24(5) \\ 
  AR1 & 27(8) & 5(0) & 22(8) \\ 
    \hline\hline
 Cause 1 Binary (sparse)  &$|S(\hat{\beta})|$  & $\#$True Positives & $\#$False Positives \\ 
 \hline
 p=20 &&\\
Independence & 10(2) & 5(0) & 5(2) \\ 
  Exchangeable & 11(2) & 5(0) & 6(2) \\ 
  AR1 & 11(2) & 5(0) & 6(2) \\ 
  \hline
   p=500 &&\\
  Independence & 17.5(6.5) & 4(1) & 13(6) \\ 
  Exchangeable & 19.5(5.5) & 4(0) & 15(5) \\ 
  AR1& 20.5(5.5) & 5(0) & 15.5(5.5) \\ 
  \hline
    p=1000 &&\\
  Independence & 17(6) & 4(0.5) & 13(6) \\ 
  Exchangeable & 20(6) & 4(1) & 16(6) \\ 
  AR1 & 21.5(5.5) & 5(0) & 17.5(5.5) \\ 
  \hline
 \end{tabular}
\end{table}

\begin{table}[!htb]
\centering
\caption{The median (MAD) number of selected variables by LASSO under the PCSH model. 
The penalty parameter is chosen using  CV+1SE.}
\label{CSH_lasso_selection2}
\small
\begin{tabular}{llll}
  \hline\hline
Cause 1 Continuous  &$|S(\hat{\beta})|$  & $\#$True Positives & $\#$False Positives \\ 
  \hline
  p=20 &&\\
  Independence & 6(1) & 5(0) & 1(1) \\ 
  Exchangeable & 6(1) & 5(0) & 1(1) \\ 
  AR1 & 6(1) & 5(0) & 1(1) \\ 
  \hline
    p=500 &&\\
 Independence & 9(2) & 5(0) & 4(3) \\ 
  Exchangeable & 12(3) & 5(0) & 7(3) \\ 
  AR1 & 10(3) & 5(0) & 5(3) \\ 
    \hline
    p=1000 &&\\
     Independence & 9(3) & 5(0) & 4(3) \\ 
  Exchangeable & 13(4) & 5(0) & 8(4) \\ 
  AR1 & 12(4) & 5(0) & 7(4) \\ 
   \hline\hline
  Cause 1 Binary (balanced)  &$|S(\hat{\beta})|$  & $\#$True Positives & $\#$False Positives \\ 
  \hline
  p=20 &&\\
Independence & 5(0) & 5(0) & 0(0) \\ 
  Exchangeable & 5(1) & 5(0) & 0(0) \\ 
  AR1 & 5(0) & 5(0) & 0(0) \\ 
  \hline
  p=500 &&\\
  Independence & 5(1) & 5(0) & 1(1) \\ 
  Exchangeable & 6(1) & 4(0) & 2(1) \\ 
  AR1 & 7(2) & 5(0) & 2(2) \\ 
  \hline
  p=1000 &&\\
  Independence & 5(1) & 5(0) & 0(0) \\ 
  Exchangeable & 6(2) & 4(0) & 1(1) \\ 
  AR1 & 6(1) & 5(0) & 1(1) \\
     \hline\hline
  Cause 1 Binary (sparse)  &$|S(\hat{\beta})|$  & $\#$True Positives & $\#$False Positives \\ 
  \hline
  p=20 &&\\
  Independence & 4(1) & 4(1) & 0(0) \\ 
  Exchangeable & 5(1) & 4(1) & 0(0) \\ 
  AR1 & 5(1) & 5(0) & 0(0) \\ 
  \hline
    p=500 &&\\
  Independence & 4(1) & 4(1) & 0(0) \\ 
  Exchangeable & 3(1) & 3(1) & 0(0) \\ 
  AR1& 4(1) & 4(1) & 0(0) \\ 
    \hline
    p=1000 &&\\
  Independence & 3(1) & 3(1) & 0(0) \\ 
  Exchangeable & 3.5(1.5) & 3(1) & 0(0) \\ 
  AR1& 4(1) & 4(1) & 0(0) \\ 
  \hline
\end{tabular}
\end{table}

\clearpage
\begin{table}[!htb]
\centering
\caption{The median (MAD) number of selected variables by LASSO under the PCSH model. 
The penalty parameter is chosen using  min AIC.} 
\label{CSH_lasso_selection3}
\small
\begin{tabular}{llll}
  \hline\hline
Cause 1 Continuous  &$|S(\hat{\beta})|$  & $\#$True Positives & $\#$False Positives \\ 
  \hline
  p=20 &&\\
Independence & 7(1) & 5(0) & 2(1) \\ 
  Exchangeable & 7(1) & 5(0) & 2(1) \\ 
  AR1 & 7(1) & 5(0) & 2(1) \\ 
  \hline
    p=500 &&\\
 Independence & 280(16.5) & 5(0) & 275(16.5) \\ 
  Exchangeable & 303.5(13) & 5(0) & 298.5(13) \\ 
  AR1 & 295.5(11.5) & 5(0) & 290.5(11.5) \\ 
    \hline
    p=1000 &&\\
      Independence & 260(15) & 5(0) & 255(15) \\ 
  Exchangeable & 292(16.5) & 5(0) & 287(16.5) \\ 
  AR1 & 282(17) & 5(0) & 277(17) \\ 
   \hline\hline
   Cause 1 Binary (balanced)  &$|S(\hat{\beta})|$  & $\#$True Positives & $\#$False Positives \\ 
  \hline
  p=20 &&\\
Independence & 7(1) & 5(0) & 2(1) \\ 
  Exchangeable & 7(1) & 5(0) & 2(1) \\ 
  AR1 & 7(1) & 5(0) & 2(1) \\ 
  \hline
  p=500 &&\\
  Independence & 364.5(16.5) & 5(0) & 359.5(16.5) \\ 
  Exchangeable & 372(17) & 5(0) & 367(17) \\ 
  AR1 & 369(15) & 5(0) & 364(15) \\ 
  \hline
  p=1000 &&\\
  Independence & 344(17.5) & 5(0) & 339.5(17.5) \\ 
  Exchangeable & 352(14.5) & 5(0) & 347(14.5) \\ 
  AR1& 362(16.5) & 5(0) & 357(16.5) \\ 
     \hline\hline
   Cause 1 Binary  (sparse) &$|S(\hat{\beta})|$  & $\#$True Positives & $\#$False Positives \\ 
     \hline
  p=20 &&\\
   Independence & 7(1) & 5(0) & 2.5(1.5) \\ 
  Exchangeable & 8(2) & 5(0) & 3(2) \\ 
  AR1 & 7(1) & 5(0) & 2(1) \\ 
    \hline
  p=500 &&\\
  Independence & 338(27.5) & 5(0) & 334(27) \\ 
  Exchangeable & 348(15.5) & 5(0) & 343(15) \\ 
  AR1 & 352.5(15.5) & 5(0) & 347.5(15.5) \\ 
    \hline
  p=1000 &&\\
  Independence & 329(27.5) & 4.5(0.5) & 324.5(27) \\ 
  Exchangeable & 336.5(19.5) & 5(0) & 331.5(19.5) \\ 
  AR1& 335.5(19.5) & 5(0) & 330.5(20) \\ 
 \hline
\end{tabular}
\end{table}
\clearpage 

\begin{table}[!htb]
\centering
\caption{The median (MAD) number of selected variables by LASSO under the PCSH model. 
The penalty parameter is chosen using   min BIC.} 
\label{CSH_lasso_selection4}
\small
\begin{tabular}{llll}
  \hline\hline
Cause 1 Continuous  &$|S(\hat{\beta})|$  & $\#$True Positives & $\#$False Positives \\ 
  \hline
  p=20 &&\\
Independence & 5(0) & 5(0) & 0(0) \\ 
  Exchangeable & 5(0) & 5(0) & 0(0) \\ 
  AR1 & 5(0) & 5(0) & 0(0) \\ 
  \hline
    p=500 &&\\
Independence & 268(17) & 5(0) & 263(17) \\ 
  Exchangeable & 291(20) & 5(0) & 286(20) \\ 
  AR1 & 286(17) & 5(0) & 281(17) \\ 
    \hline
    p=1000 &&\\
     Independence & 253(12.5) & 5(0) & 248(12.5) \\ 
  Exchangeable & 268.5(17) & 5(0) & 263.5(17) \\ 
  AR1 & 265(15) & 5(0) & 260(15) \\ 
\hline\hline
  Cause 1 Binary (balanced)  &$|S(\hat{\beta})|$  & $\#$True Positives & $\#$False Positives \\ 
  \hline
  p=20 &&\\
 Independence & 5(0) & 5(0) & 0(0) \\ 
  Exchangeable & 5(0) & 5(0) & 0(0) \\ 
  AR1 & 5(0) & 5(0) & 0(0) \\ 
  \hline
  p=500 &&\\
  Independence & 354.5(17.5) & 5(0) & 349.5(17) \\ 
  Exchangeable & 359(23) & 5(0) & 354(23) \\ 
  AR1 & 359(16) & 5(0) & 354(16) \\ 
  \hline
  p=1000 &&\\
  Independence & 341(17) & 5(0) & 336(17) \\ 
  Exchangeable & 345.5(16) & 5(0) & 340.5(16) \\ 
  AR1 & 354.5(18.5) & 5(0) & 349.5(18.5) \\ 
  \hline\hline
  Cause 1 Binary (sparse)  &$|S(\hat{\beta})|$  & $\#$True Positives & $\#$False Positives \\ 
   \hline
  p=20 &&\\
  Independence & 5(1) & 4(0) & 0(0) \\ 
  Exchangeable & 5(1) & 4(0) & 0(0) \\ 
  AR1 & 5(0) & 5(0) & 0(0) \\ 
   \hline
  p=500 &&\\
  Independence & 319(41) & 5(0) & 315(40.5) \\ 
  Exchangeable & 345(18) & 5(0) & 340(18) \\ 
  AR1 & 343(23) & 5(0) & 338(23) \\ 
   \hline
  p=1000 &&\\
  Independence & 321.5(26) & 5(0) & 317(26) \\ 
  Exchangeable & 323(24.5) & 5(0) & 318(24) \\ 
  AR1 & 322.5(24) & 5(0) & 318(24) \\
  \hline
\end{tabular}
\end{table}

\begin{table}[!htb]
\centering
\caption{The median (MAD) number of selected variables by LASSO under the PCSH model. 
The penalty parameter is chosen using  elbow AIC.} 
\label{CSH_lasso_selection5}
\small
\begin{tabular}{llll}
  \hline\hline
Cause 1 Continuous  &$|S(\hat{\beta})|$  & $\#$True Positives & $\#$False Positives \\ 
  \hline
  p=20 &&\\
Independence & 5.5(0.5) & 5(0) & 0.5(0.5) \\ 
  Exchangeable & 5(1) & 5(0) & 0(0) \\ 
  AR1 & 5(1) & 5(0) & 0(0) \\ 
  \hline
    p=500 &&\\
Independence & 225(21) & 5(0) & 220(21) \\ 
  Exchangeable& 27.5(18.5) & 5(0) & 22.5(18.5) \\ 
  AR1 & 200(51) & 5(0) & 195(51) \\ 
    \hline
    p=1000 &&\\
 Independence & 162.5(72) & 5(0) & 157.5(72) \\ 
  Exchangeable & 31.5(22) & 5(0) & 26.5(22) \\ 
  AR1 & 61(49) & 5(0) & 56(49) \\ 
\hline\hline
  Cause 1 Binary (balanced)  &$|S(\hat{\beta})|$  & $\#$True Positives & $\#$False Positives \\ 
  \hline
  p=20 &&\\
Independence & 6(1) & 5(0) & 1(1) \\ 
  Exchangeable & 5(1) & 5(0) & 0(0) \\ 
  AR1 & 5(1) & 5(0) & 0(0) \\ 
  \hline
  p=500 &&\\
  Independence & 271.5(41) & 5(0) & 266.5(41) \\ 
  Exchangeable & 171.5(127.5) & 5(0) & 166.5(127.5) \\ 
  AR1& 277(40) & 5(0) & 272(40) \\ 
  \hline
  p=1000 &&\\ 
  Independence & 271(26) & 5(0) & 266(26) \\ 
  Exchangeable & 70.5(60) & 5(0) & 65.5(59.5) \\ 
  AR1 & 247(68) & 5(0) & 242(68) \\ 
\hline\hline
 Cause 1 Binary (sparse) &$|S(\hat{\beta})|$  & $\#$True Positives & $\#$False Positives \\ 
   \hline
  p=20 &&\\
 Independence & 5(1) & 4(1) & 1(1) \\ 
  Exchangeable & 5(1) & 5(0) & 1(1) \\ 
  AR1 & 6(1) & 5(0) & 1(1) \\ 
     \hline
  p=500 &&\\
  Independence & 222.5(77.5) & 5(0) & 217.5(77) \\ 
  Exchangeable & 105.5(79.5) & 5(0) & 100.5(78.5) \\ 
  AR1 & 256(40) & 5(0) & 251(40) \\ 
     \hline
  p=1000 &&\\
  Independence & 114.5(75.5) & 4(1) & 110.5(75.5) \\ 
  Exchangeable & 101(68.5) & 5(0) & 96(68) \\ 
  AR1 & 210(73) & 5(0) & 205(73) \\ 
  \hline
\end{tabular}
\end{table}

 \begin{table}[!htb]
\centering
\caption{The median (MAD) number of selected variables by LASSO under the PCSH model. 
The penalty parameter is chosen using  elbow BIC.} 
\label{CSH_lasso_selection6}
\small
\begin{tabular}{llll}
  \hline\hline
Cause 1 Continuous  &$|S(\hat{\beta})|$  & $\#$True Positives & $\#$False Positives \\ 
  \hline
  p=20 &&\\
Independence & 4(0) & 4(0) & 0(0) \\ 
  Exchangeable & 4(0) & 4(0) & 0(0) \\ 
  AR1 & 4(0) & 4(0) & 0(0) \\
  \hline
    p=500 &&\\
    Independence & 5(1) & 5(0) & 0(0) \\ 
  Exchangeable & 5(1) & 4(1) & 0(0) \\ 
  AR1 & 5(1) & 5(0) & 0(0) \\ 
    \hline
    p=1000 &&\\
 Independence & 6.5(1.5) & 5(0) & 1.5(1.5) \\ 
  Exchangeable & 5(1) & 4(1) & 0(0) \\ 
  AR1& 6(1) & 5(0) & 1(1) \\ 
    \hline\hline
    Cause 1 Binary (balanced)  &$|S(\hat{\beta})|$  & $\#$True Positives & $\#$False Positives \\ 
  \hline
  p=20 &&\\
  Independence & 4(0) & 4(0) & 0(0) \\ 
  Exchangeable & 4(0) & 4(0) & 0(0) \\ 
  AR1 & 4(0) & 4(0) & 0(0) \\ 
  \hline
  p=500 &&\\
  Independence & 6(1) & 5(0) & 1(1) \\ 
  Exchangeable & 5(1) & 4(0) & 1(1) \\ 
  AR1 & 6(1) & 5(0) & 1(1) \\ 
  \hline
  p=1000 &&\\
  Independence & 8(2) & 5(0) & 3(2) \\ 
  Exchangeable & 6(2) & 4(0) & 2(1) \\ 
  AR1 & 7(2) & 5(0) & 2(1) \\ 
   \hline\hline
    Cause 1 Binary  (sparse) &$|S(\hat{\beta})|$  & $\#$True Positives & $\#$False Positives \\ 
      \hline
  p=20 &&\\
Independence & 3(1) & 3(1) & 0(0) \\ 
  Exchangeable & 3(1) & 3(0.5) & 0(0) \\ 
  AR1 & 4(1) & 4(0.5) & 0(0) \\ 
    \hline
  p=500 &&\\
  Independence & 6(2) & 4(1) & 2(2) \\ 
  Exchangeable & 5(1) & 4(0) & 2(1) \\ 
  AR1 & 6(1) & 4(1) & 2(1) \\ 
    \hline
  p=1000 &&\\
  Independence & 7(2) & 4(1) & 3(2) \\ 
  Exchangeable & 7(2) & 4(1) & 3(2) \\ 
  AR1 & 7(2) & 4(0) & 3(2) \\ 
  \hline
\end{tabular}
\end{table}


\clearpage
\begin{table}[!htb]
\centering
\caption{The median (MAD) number of selected variables by LASSO under the PSDH model. 
The number of steps is chosen using  CV10.}
\label{FG_boost_var1}
\small
 \begin{tabular}{llll}
  \hline\hline
Cause 1 Continuous  & $|S(\hat{\beta})|$ & $\#$True Positives & $\#$False Positives \\ 
\hline
p=20 & & \\
Independence & 7(2) & 5(0) & 2(2) \\ 
  Exchangeable & 10(3) & 5(0) & 5(3) \\ 
  AR1 & 9(2) & 5(0) & 4(2) \\
   \hline
 p=500 & & \\ 
  Independence & 5(0) & 5(0) & 0(0) \\ 
  Exchangeable & 6(1) & 5(0) & 1(1) \\ 
  AR1 & 6(1) & 5(0) & 1(1) \\
   \hline
 p=1000 & & \\ 
  Independence & 5(0) & 5(0) & 0(0) \\ 
  Exchangeable & 5(0) & 5(0) & 0(0) \\ 
  AR1 & 5(0) & 5(0) & 0(0) \\ 
   \hline\hline
  Cause 1 Binary (balanced)   & $|S(\hat{\beta})|$ & $\#$True Positives & $\#$False Positives \\ 
\hline
p=20 & & \\
Independence & 6(1) & 5(0) & 1(1) \\ 
  Exchangeable & 7(2) & 5(0) & 2(2) \\ 
  AR1 & 6(1) & 5(0) & 1(1) \\ 
   \hline
 p=500 & & \\ 
  Independence & 5(1) & 5(0) & 0(0) \\ 
  Exchangeable & 4(1) & 4(0) & 0(0) \\ 
  AR1 & 4.5(0.5) & 4(0) & 0(0) \\ 
   \hline
 p=1000 & & \\ 
  Independence & 5(1) & 4(1) & 0(0) \\ 
  Exchangeable & 4(0) & 4(0) & 0(0) \\ 
  AR1 & 4(0) & 4(0) & 0(0) \\ 
   \hline\hline
     Cause 1 Binary (sparse)  & $|S(\hat{\beta})|$ & $\#$True Positives & $\#$False Positives \\ 
     \hline
  p=20 & & \\
  Independence & 5(1) & 4(1) & 0(0) \\ 
  Exchangeable & 6(1) & 5(0) & 1(1) \\ 
  AR1 & 5(1) & 5(0) & 0(0) \\ 
    \hline
   p=500 & & \\
  Independence & 3(1) & 2(1) & 0(0) \\ 
  Exchangeable & 3(0) & 3(0) & 0(0) \\ 
  AR1 & 3(0) & 3(0) & 0(0) \\ 
      \hline
   p=1000 & & \\
  Independence & 2(1) & 2(1) & 0(0) \\ 
  Exchangeable & 3(0) & 3(0) & 0(0) \\ 
  AR1 & 3(0) & 3(0) & 0(0) \\ 
  \hline
\end{tabular}
\end{table}

\begin{table}[!htb]
\centering
\caption{The median (MAD) number of selected variables by LASSO under the PSDH model. 
The number of steps is chosen using  min AIC.}
\label{FG_boost_var2}
\small
 \begin{tabular}{llll}
  \hline\hline
Cause 1 Continuous  & $|S(\hat{\beta})|$ & $\#$True Positives & $\#$False Positives \\ 
\hline
p=20 & & \\
Independence & 7(1) & 5(0) & 2(1) \\ 
  Exchangeable & 7(1) & 5(0) & 2(1) \\ 
  AR1 & 7(1) & 5(0) & 2(1) \\
   \hline
 p=500 & & \\ 
  Independence & 7(1) & 5(0) & 2(1) \\ 
  Exchangeable & 8(1) & 5(0) & 3(1) \\ 
  AR1 & 7.5(1.5) & 5(0) & 2.5(1.5) \\
   \hline
 p=1000 & & \\ 
  Independence & 7(1) & 5(0) & 2(1) \\ 
  Exchangeable & 7(1) & 5(0) & 2(1) \\ 
  AR1 & 7(1) & 5(0) & 2(1) \\ 
   \hline\hline
   Cause 1 Binary (balanced)  & $|S(\hat{\beta})|$ & $\#$True Positives & $\#$False Positives \\ 
\hline
p=20 & & \\
Independence & 7(1) & 5(0) & 2(1) \\ 
  Exchangeable & 7(1) & 5(0) & 2.5(1.5) \\ 
  AR1 & 7(1) & 5(0) & 2(1) \\ 
  \hline
  p=500 & & \\
  Independence & 7(1) & 5(0) & 2(0) \\ 
  Exchangeable & 7(1) & 5(0) & 2(1) \\ 
  AR1 & 7(0) & 5(0) & 2(0) \\ 
  \hline
  p=1000 & & \\
  Independence & 7(1) & 5(0) & 2(1) \\ 
  Exchangeable & 6(1) & 5(0) & 2(1) \\ 
  AR1 & 7(0) & 5(0) & 2(0) \\ 
      \hline\hline
   Cause 1 Binary (sparse) & $|S(\hat{\beta})|$ & $\#$True Positives & $\#$False Positives \\ 
   \hline
p=20 & & \\
   Independence & 6(1) & 5(0) & 2(1) \\ 
  Exchangeable & 7(1) & 5(0) & 2(1) \\ 
  AR1 & 6(1) & 5(0) & 2(1) \\ 
  \hline
p=500 & & \\
  Independence & 5(1) & 4(1) & 1(1) \\ 
  Exchangeable & 6(1) & 4(0) & 2(0.5) \\ 
  AR1 & 6(0) & 5(0) & 2(1) \\ 
  \hline
p=1000 & & \\
  Independence & 4(1) & 3(1) & 1(1) \\ 
  Exchangeable & 6(0) & 4(0) & 2(0) \\ 
  AR1 & 6(0) & 4(0) & 2(1) \\ 
\hline
\end{tabular}
\end{table}

\begin{table}[!htb]
\centering
\small
\caption{The median (MAD) number of selected variables by LASSO under the PSDH model. 
The number of steps is chosen using  min BIC.}
\label{FG_boost_var3}
 \begin{tabular}{llll}
  \hline\hline
Cause 1 Continuous  & $|S(\hat{\beta})|$ & $\#$True Positives & $\#$False Positives \\ 
\hline
p=20 & & \\
Independence & 5(0) & 5(0) & 0(0) \\ 
  Exchangeable & 5(0) & 5(0) & 0(0) \\ 
  AR1 & 5(0) & 5(0) & 0(0) \\ 
   \hline
 p=500 & & \\ 
  Independence & 7(1) & 5(0) & 2(1) \\ 
  Exchangeable & 7(1) & 5(0) & 2(1) \\ 
  AR1 & 7(1) & 5(0) & 2(1) \\
   \hline
 p=1000 & & \\ 
  Independence & 7(1) & 5(0) & 2(1) \\ 
  Exchangeable & 7(1) & 5(0) & 2(1) \\ 
  AR1 & 7(1) & 5(0) & 2(1) \\ 
   \hline\hline
   Cause 1 Binary (balanced)  & $|S(\hat{\beta})|$ & $\#$True Positives & $\#$False Positives \\ 
\hline
p=20 & & \\
Independence & 5(0) & 5(0) & 0(0) \\ 
  Exchangeable & 5(0) & 5(0) & 0(0) \\ 
  AR1 & 5(0) & 5(0) & 0(0) \\ 
  \hline
  p=500 & & \\
  Independence & 7(1) & 5(0) & 2(0.5) \\ 
  Exchangeable & 6(1) & 5(0) & 2(1) \\ 
  AR1 & 7(0.5) & 5(0) & 2(0) \\ 
  \hline
  p=1000 & & \\
  Independence & 7(1) & 5(0) & 2(1) \\ 
  Exchangeable  & 6(1) & 5(0) & 2(1) \\ 
  AR1 & 7(0) & 5(0) & 2(0) \\ 
      \hline\hline
   Cause 1 Binary (sparse) & $|S(\hat{\beta})|$ & $\#$True Positives & $\#$False Positives \\ 
   \hline
p=20 & & \\
   Independence & 4(1) & 4(1) & 0(0) \\ 
  Exchangeable & 5(0) & 5(0) & 0(0) \\ 
  AR1 & 5(0) & 5(0) & 0(0) \\ 
  \hline
p=500 & & \\
  Independence. & 4.5(0.5) & 4(1) & 1(1) \\ 
  Exchangeable. & 6(1) & 4(0) & 2(1) \\ 
  AR1 & 6(0.5) & 4.5(0.5) & 2(1) \\ 
  \hline
p=1000 & & \\
  Independence & 4(1) & 3(1) & 1(1) \\ 
  Exchangeable & 6(0) & 4(0) & 2(0) \\ 
  AR1 & 6(0) & 4(0) & 2(1) \\ 
   \hline
\end{tabular}
\end{table}

\begin{table}[!htb]
\centering
\small
\caption{The median (MAD) number of selected variables by LASSO under the PSDH model. 
The number of steps is chosen using   elbow AIC.}
\label{FG_boost_var4}
 \begin{tabular}{llll}
  \hline\hline
Cause 1 Continuous  & $|S(\hat{\beta})|$ & $\#$True Positives & $\#$False Positives \\ 
\hline
p=20 & & \\
Independence & 3(0) & 3(0) & 0(0) \\ 
  Exchangeable & 4(1) & 4(1) & 0(0) \\ 
  AR1 & 4.5(0.5) & 4.5(0.5) & 0(0) \\ 
   \hline
 p=500 & & \\ 
  Independence & 3(0) & 3(0) & 0(0) \\ 
  Exchangeable & 4(0) & 4(0) & 0(0) \\ 
  AR1 & 4(1) & 4(1) & 0(0) \\
   \hline
 p=1000 & & \\ 
  Independence & 4(1) & 4(1) & 0(0) \\ 
  Exchangeable & 4(1) & 4(1) & 0(0) \\ 
  AR1 & 4.5(0.5) & 4.5(0.5) & 0(0) \\ 
   \hline\hline
   Cause 1 Binary (balanced)  & $|S(\hat{\beta})|$ & $\#$True Positives & $\#$False Positives \\ 
\hline
p=20 & & \\
Independence & 3(0) & 3(0) & 0(0) \\ 
  Exchangeable & 3.5(0.5) & 3.5(0.5) & 0(0) \\ 
  AR1 & 4(0) & 4(0) & 0(0) \\ 
  \hline
p=500 & & \\
  Independence & 3(0) & 3(0) & 0(0) \\ 
  Exchangeable & 3(1) & 3(1) & 0(0) \\ 
  AR1 & 4(0) & 4(0) & 0(0) \\ 
  \hline
p=1000 & & \\
  Independence & 3(0) & 3(0) & 0(0) \\ 
  Exchangeable & 3(1) & 3(1) & 0(0) \\ 
  AR1 & 4(0) & 4(0) & 0(0) \\ 
   \hline\hline
   Cause 1 Binary (sparse) & $|S(\hat{\beta})|$ & $\#$True Positives & $\#$False Positives \\ 
   \hline
p=20 & & \\
   Independence & 3(1) & 3(1) & 0(0) \\ 
  Exchangeable & 3(1) & 3(1) & 0(0) \\ 
  AR1 & 4(0.5) & 4(0.5) & 0(0) \\ 
     \hline
p=500 & & \\
  Independence & 3(1) & 3(1) & 0(0) \\ 
  Exchangeable & 3(1) & 3(1) & 0(0) \\ 
  AR1 & 4(0.5) & 4(0.5) & 0(0) \\ 
     \hline
p=1000 & & \\
  Independence & 2(1) & 2(1) & 0(0) \\ 
  Exchangeable & 3(1) & 3(1) & 0(0) \\ 
  AR1 & 4(0) & 4(0) & 0(0) \\ 
  \hline
\end{tabular}
\end{table}

\begin{table}[!htb]
\centering
\small
\caption{The median (MAD) number of selected variables by LASSO under the PSDH model. 
The number of steps is chosen using   elbow BIC.}
\label{FG_boost_var5}
 \begin{tabular}{llll}
  \hline\hline
Cause 1 Continuous  & $|S(\hat{\beta})|$ & $\#$True Positives & $\#$False Positives \\ 
\hline
p=20 & & \\
Independence & 3(0) & 3(0) & 0(0) \\ 
  Exchangeable & 4(1) & 4(1) & 0(0) \\ 
  AR1 & 4.5(0.5) & 4.5(0.5) & 0(0) \\ 
   \hline
 p=500 & & \\ 
  Independence & 3(0) & 3(0) & 0(0) \\ 
  Exchangeable & 4(0) & 4(0) & 0(0) \\ 
  AR1 & 4(1) & 4(1) & 0(0) \\
   \hline
 p=1000 & & \\ 
  Independence & 4(1) & 4(1) & 0(0) \\ 
  Exchangeable & 4(1) & 4(1) & 0(0) \\ 
  AR1 & 4.5(0.5) & 4.5(0.5) & 0(0) \\ 
   \hline\hline
   Cause 1 Binary (balanced)  & $|S(\hat{\beta})|$ & $\#$True Positives & $\#$False Positives \\ 
\hline
p=20 & & \\
Independence & 3(0) & 3(0) & 0(0) \\ 
  Exchangeable & 3.5(0.5) & 3.5(0.5) & 0(0) \\ 
  AR1 & 4(0) & 4(0) & 0(0) \\ 
  \hline
p=500 & & \\
  Independence & 3(0) & 3(0) & 0(0) \\ 
  Exchangeable & 3(1) & 3(1) & 0(0) \\ 
  AR1 & 4(0) & 4(0) & 0(0) \\ 
  \hline
p=1000 & & \\
  Independence & 3(0) & 3(0) & 0(0) \\ 
  Exchangeable & 3(1) & 3(1) & 0(0) \\ 
  AR1& 4(0) & 4(0) & 0(0) \\ 
   \hline\hline
   Cause 1 Binary (sparse) & $|S(\hat{\beta})|$ & $\#$True Positives & $\#$False Positives \\ 
     \hline
p=20 & & \\
   Independence & 3(1) & 3(1) & 0(0) \\ 
  Exchangeable & 3(1) & 3(1) & 0(0) \\ 
  AR1 & 4(0.5) & 4(0.5) & 0(0) \\ 
       \hline
p=500 & & \\
  Independence & 3(1) & 3(1) & 0(0) \\ 
  Exchangeable & 3(1) & 3(1) & 0(0) \\ 
  AR1 & 4(0.5) & 4(0.5) & 0(0) \\ 
       \hline
p=1000 & & \\
  Independence & 2(1) & 2(1) & 0(0) \\ 
  Exchangeable & 3(1) & 3(1) & 0(0) \\ 
  AR1 & 4(0) & 4(0) & 0(0) \\ 
   \hline
\end{tabular}
\end{table}

\clearpage

The following two figures correspond to the main paper Figure 2 and 3, where `density()' failed in some cases to properly depict the empirical distribution of $\widehat{\cif}_1(2)$.

 \begin{figure}[!htb]
  \caption{Boxplots of $\widehat{\cif}_1(2)$, estimated under the PCSH model with LASSO, for balanced binary covariates. The three columns correspond to $p = 20$, 500, and 1000. The rows correspond to different ways of selecting $\lambda$, from top to botton: 1) CV10, 2) CV+1SE, 3) minimum AIC, 4) minimum BIC, 5) elbow AIC and 6) elbow BIC. The true $\cif_1(2 | z_{0})=0.11$. 
} 
\label{CSH_LASSO_CIF1_2yr_binary}
  \centering
    \includegraphics[width=12cm]{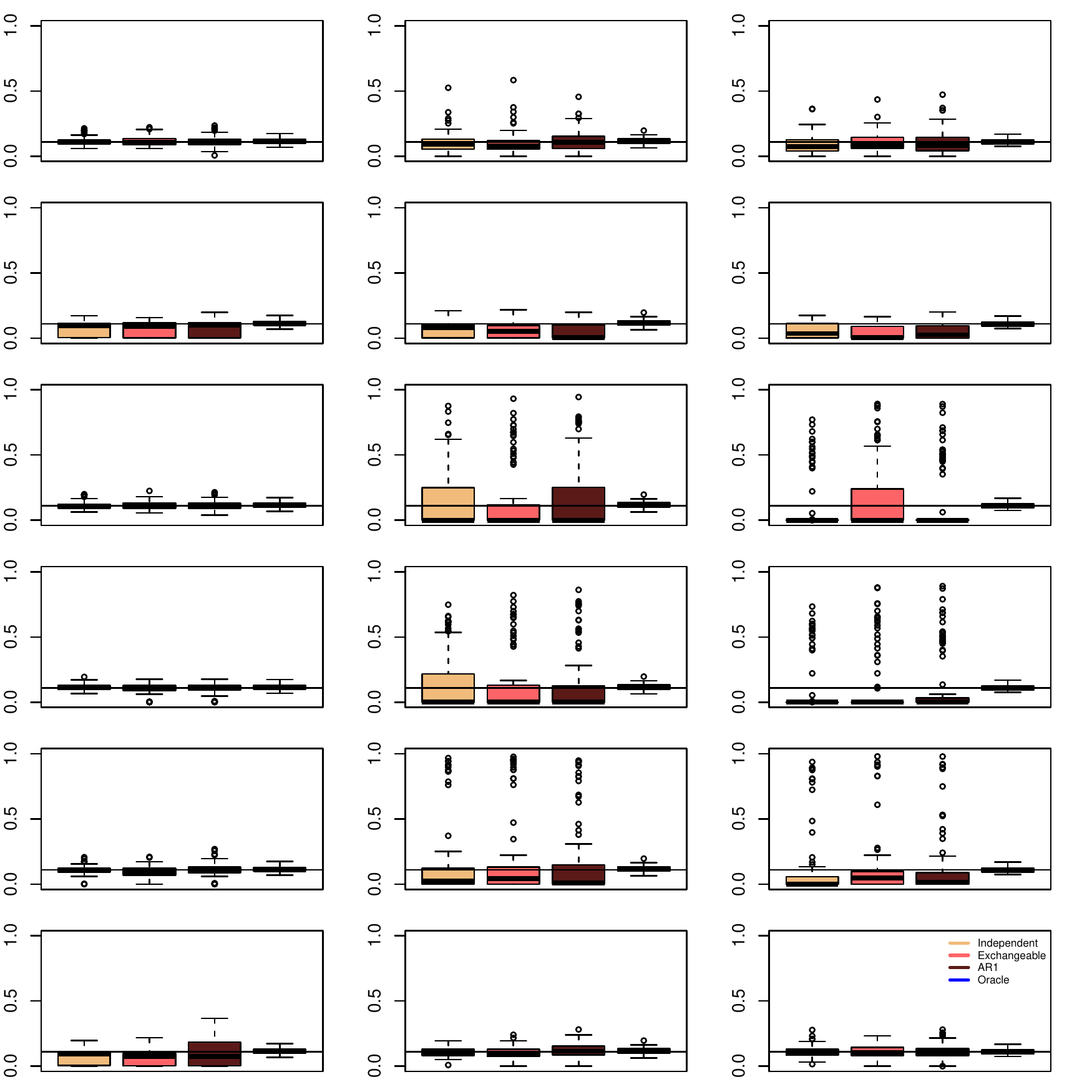}
         \end{figure}
   
 \begin{figure}[!htb]
  \caption{Boxplots of $\widehat{\cif}_1(2)$, estimated under the PCSH model with LASSO, for sparse binary covariates. The three columns correspond to $p = 20$, 500, and 1000. The rows correspond to different ways of selecting $\lambda$, from top to botton: 1) CV10, 2) CV+1SE, 3) minimum AIC, 4) minimum BIC, 5) elbow AIC and 6) elbow BIC. The true $\cif_1(2 | z_{0})=0.11$. 
} 
\label{CSH_LASSO_CIF1_2yr_sparse_binary}
  \centering
    \includegraphics[width=12cm]{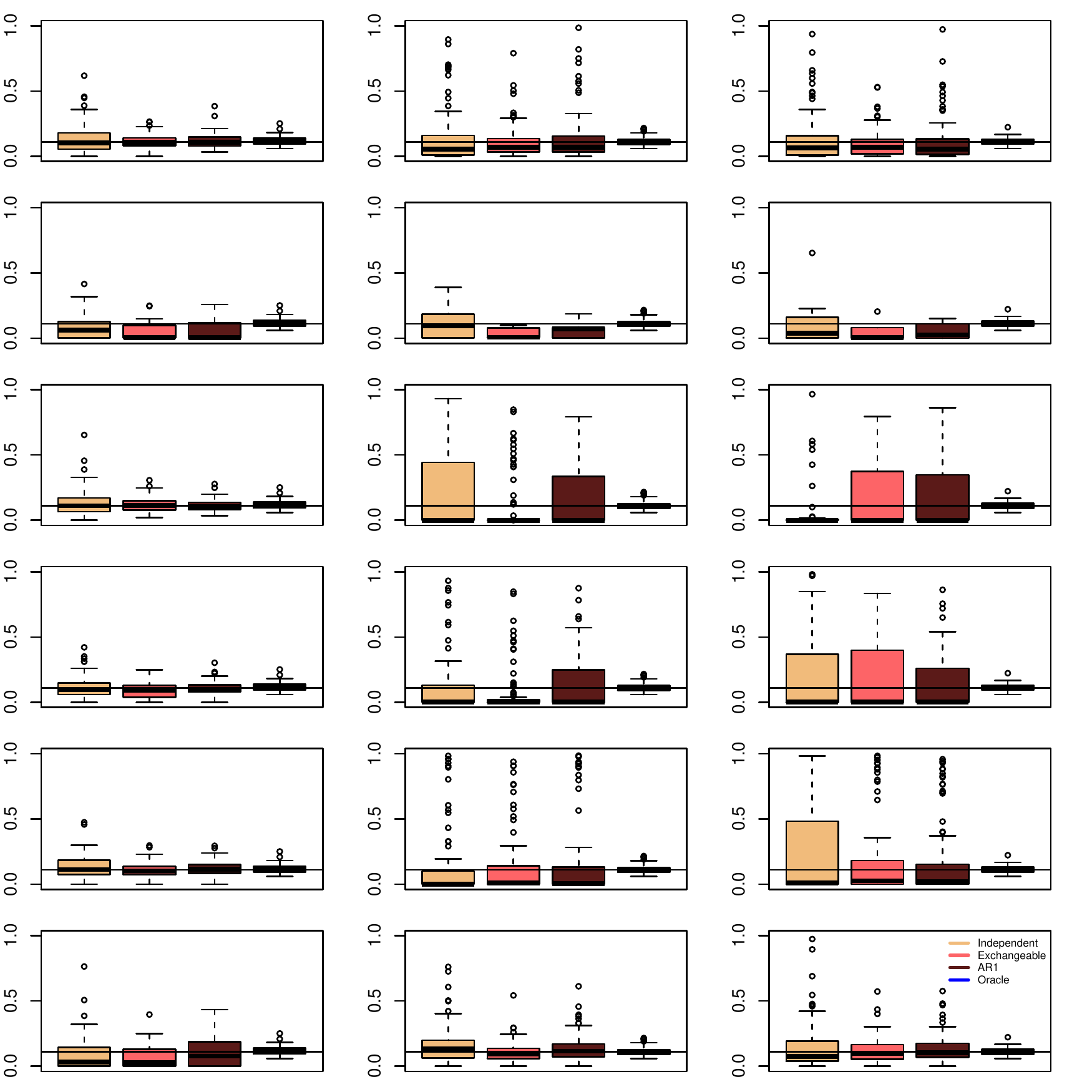}
         \end{figure}
   
 \begin{figure}[!htb]
 \caption{Predicted 16-group cumulative incidence functions for non-cancer and cancer mortality, based on 4 non-cancer and 4 cancer strata under the PCSH model.}
\label{noncancer_group16}
  \centering
\includegraphics[width=6cm]{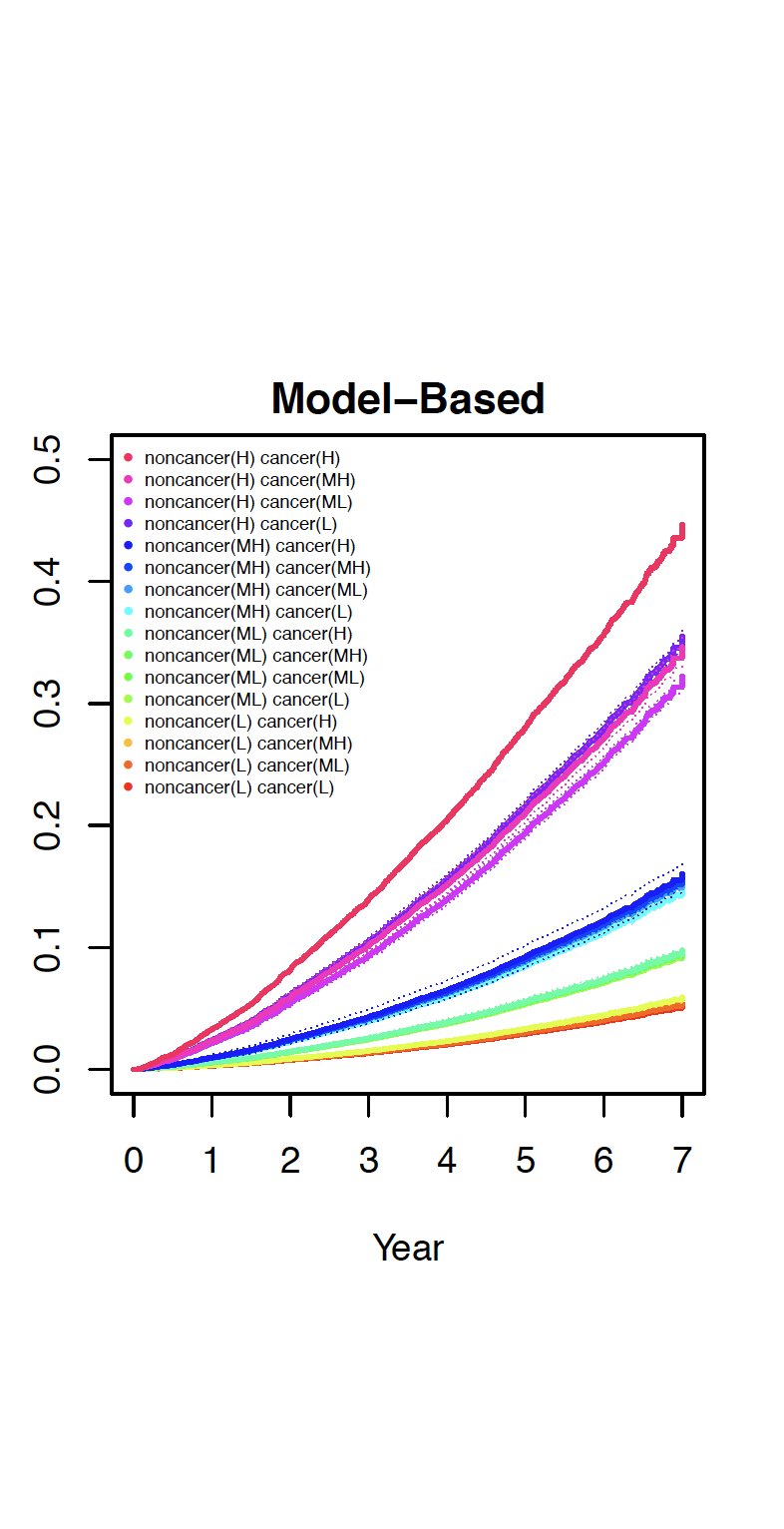}
\includegraphics[width=6.2cm]{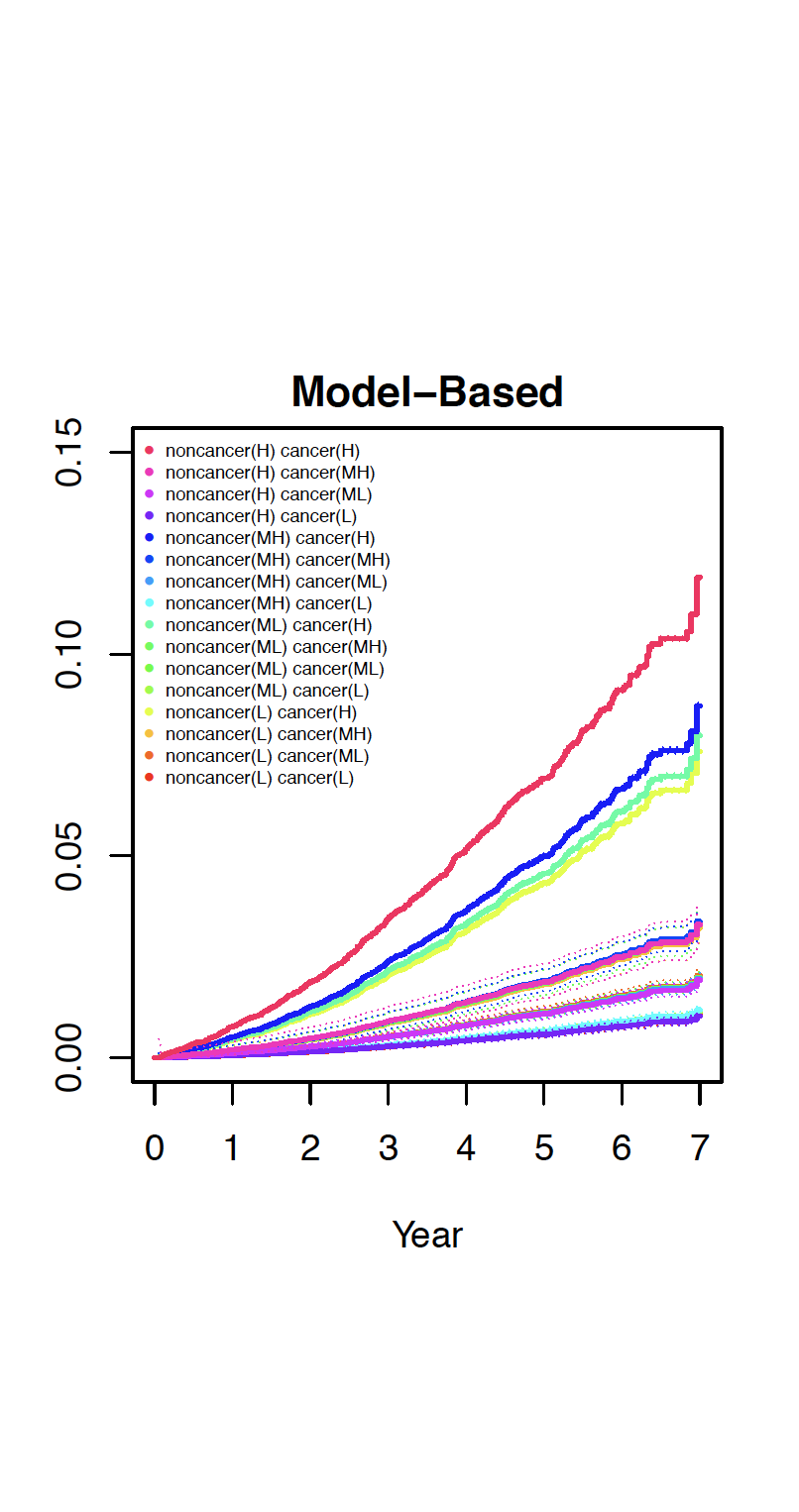}
    \end{figure}

\end{document}